\documentclass[preprint]{aastex63}

\usepackage{amsmath}

\usepackage{nicefrac}

\usepackage{tikz}
\usetikzlibrary{bayesnet}

\renewcommand{\arg}[1]{\! \left( #1 \right)}
\newcommand{\args}[1]{\big( #1 \big)}
\newcommand{\pcond}[2]{p \arg{ #1 \mid #2 }}
\newcommand{\pconds}[2]{p \args{ #1 \mid #2 }}

\newcommand\teff{T_{\mathrm{eff}}}
\newcommand\feh{\left[ \mathrm{Fe} / \mathrm{H} \right]}
\newcommand\logg{\log g}

\received{30 June 2020}
\revised{18 November 2020}
\accepted{7 December 2020}
\submitjournal{ApJ}

\shorttitle{Data-Driven Stellar Models}
\shortauthors{Green et al.}

\begin{document}

\title{Data-Driven Stellar Models}

\correspondingauthor{Gregory M. Green}
\email{gregorymgreen@gmail.com}

\author[0000-0001-5417-2260]{Gregory M. Green}
\affiliation{Max Planck Institute for Astronomy \\
K\"{o}nigstuhl 17 \\
D-69117 Heidelberg, Germany}

\author[0000-0003-4996-9069]{Hans-Walter Rix}
\affiliation{Max Planck Institute for Astronomy \\
K\"{o}nigstuhl 17 \\
D-69117 Heidelberg, Germany}

\author[0000-0002-2971-5208]{Leon Tschesche}
\affiliation{Max Planck Institute for Astronomy \\
K\"{o}nigstuhl 17 \\
D-69117 Heidelberg, Germany}

\author[0000-0003-2808-275X]{Douglas Finkbeiner}
\affiliation{Harvard Astronomy, Harvard-Smithsonian Center for Astrophysics \\
60 Garden St. \\
Cambridge, MA 02138, USA}

\author[0000-0002-2250-730X]{Catherine Zucker}
\affiliation{Harvard Astronomy, Harvard-Smithsonian Center for Astrophysics \\
60 Garden St. \\
Cambridge, MA 02138, USA}

\author[0000-0002-3569-7421]{Edward F. Schlafly}
\affiliation{Lawrence Livermore National Laboratory \\
7000 East Ave. \\
Livermore, CA 94550, USA}

\author[0000-0002-0993-6089]{Jan Rybizki}
\affiliation{Max Planck Institute for Astronomy \\
K\"{o}nigstuhl 17 \\
D-69117 Heidelberg, Germany}

\author[0000-0001-9256-5516]{Morgan Fouesneau}
\affiliation{Max Planck Institute for Astronomy \\
K\"{o}nigstuhl 17 \\
D-69117 Heidelberg, Germany}

\author[0000-0002-6274-6612]{Ren\'{e} Andrae}
\affiliation{Max Planck Institute for Astronomy \\
K\"{o}nigstuhl 17 \\
D-69117 Heidelberg, Germany}

\author[0000-0003-2573-9832]{Joshua Speagle}
\affiliation{Harvard Astronomy, Harvard-Smithsonian Center for Astrophysics \\
60 Garden St. \\
Cambridge, MA 02138, USA}

\begin{abstract}

We develop a data-driven model to map stellar parameters ($\teff$, $\logg$, and $\feh$) accurately and precisely to broad-band stellar photometry. This model must, and does, simultaneously constrain the passband-specific dust reddening vector in the Milky Way, $\vec{R}$. The model uses a neural network to learn the (de-reddened) absolute magnitude in one band and colors across many bands, given stellar parameters from spectroscopic surveys and parallax constraints from Gaia. To demonstrate the effectiveness of this approach, we train our model on a dataset with spectroscopic parameters from LAMOST, APOGEE and GALAH, Gaia parallaxes, and optical and near-infrared photometry from Gaia, Pan-STARRS~1, 2MASS and WISE. Testing the model on these datasets leads to an excellent fit and a precise -- and by construction accurate -- prediction of the color-magnitude diagrams in many bands. This flexible approach rigorously links spectroscopic and photometric surveys, and also results in an improved, $\teff$-dependent $\vec{R}$. As such, it provides a simple and accurate method for predicting photometry in stellar evolutionary models. Our model will form a basis to infer stellar properties, distances and dust extinction from photometric data, which should be of great use in 3D mapping of the Milky Way. Our trained model may be obtained at \href{https://doi.org/10.5281/zenodo.3902382}{DOI:10.5281/zenodo.3902382}.

\end{abstract}

\keywords{Astrostatistics (1882), Neural networks (1933), Stellar photometry (1620), Interstellar dust extinction (837)}


\section{Introduction} \label{sec:intro}

Much in astronomy relies on accurate determination and knowledge of stellar properties. The most accurate methods for determining stellar properties from observables are spectroscopic modeling and astroseismology. However, spectra and precise time-series photometry are unavailable for most objects. Broad-band photometry, while providing coarser-grained information about stellar properties, is available in much greater abundance. To good approximation, stars can be characterized by $\vec{\theta} = \bigl (\teff, \logg, \feh \bigr )$. To constrain stellar parameters with broad-band photometry $\vec{m}$ requires a forward model,
\begin{align}
    f \, : \, \vec{\theta} \, \mapsto \vec{M} \, 
\end{align}
that maps stellar parameters to absolute magnitude $\vec{M}$ in the observed passbands.

Theoretical stellar models provide one way of mapping from stellar parameters to absolute magnitudes. Beginning with initial mass, age and abundances, one can use a stellar evolutionary model to predict the bolometric luminosity and radius (which map into the photospheric $\teff$ and $\logg$), along with the atmospheric abundances. Synthetic stellar atmospheric models then map these parameters to broad-band photometric magnitudes. These predictions can then be used to infer stellar parameters, as done for example by the \texttt{StarHorse} code \citep{Queiroz2018,Santiago2016}. A number of systematic effects can, and do, affect these theoretically derived absolute magnitudes, including poorly modeled molecular lines and microturbulence in low-temperature stars, inaccurate instrumental and atmospheric transmission curves, and errors in photometric zero-point calibrations \citep{Casagrande2014,Casagrande2018}. One method of dealing with these inaccuracies in theoretical models is to apply empirical corrections based on the observed photometry of stars of known type.

In this work, we cut out the middle-man, and learn the mapping from stellar parameters $\vec{\theta}$ to absolute magnitudes directly from the data, without explicit reference to theoretical models. This method should be simpler, more rigorous and quicker to adapt to new photometric datasets than the more traditional approach of tweaking theoretical models to better match the data.

We derive an empirical mapping from stellar atmospheric parameters to absolute magnitudes, using stars with well-measured parallaxes and $\vec{\theta} = \left(\teff, \logg, \feh \right)$ determined by spectroscopy (hereafter referred to as ``spectroscopic features''). This method can be applied to derive models of any combination of photometric passbands, and can be expanded to include dependence on more detailed element abundances (such as $\left[ \alpha / \mathrm{Fe} \right]$). Our method represents the function ${f \, : \, \vec{\theta} \, \mapsto \vec{M}}$ using a simple neural network architecture. At the same time, we must and do learn a simple model of dust extinction in the chosen photometric passbands. Because our mapping from spectroscopic features to photometry is learned directly from observations, the resulting stellar colors are more accurate -- by construction -- than those produced by theoretical models. This should enable more reliable determination of stellar parameters from broadband photometry, allowing the community to better leverage large photometric surveys.

We demonstrate the performance of this approach with a dataset consisting of stars with Gaia parallaxes, optical and near-infrared photometry from Gaia, Pan-STARRS~1, 2MASS and WISE, and spectroscopic type determinations from LAMOST, APOGEE and GALAH.

The paper is organized as follows. In \S\ref{sec:method}, we lay out our machine-learning method in a manner that is agnostic to the specific photometric, parallax and spectroscopic datasets used. In \S\ref{sec:data}, we discuss the composition of our input dataset, including spectroscopic features, parallaxes, photometry and reddening estimates. In \S\ref{sec:training}, we discuss the training procedure for our model. In \S\ref{sec:results}, we present our trained model, which we validate in \S\ref{sec:validation}. \S\ref{sec:discussion} discusses possible uses and extensions of our approach.

\section{Method}
\label{sec:method}

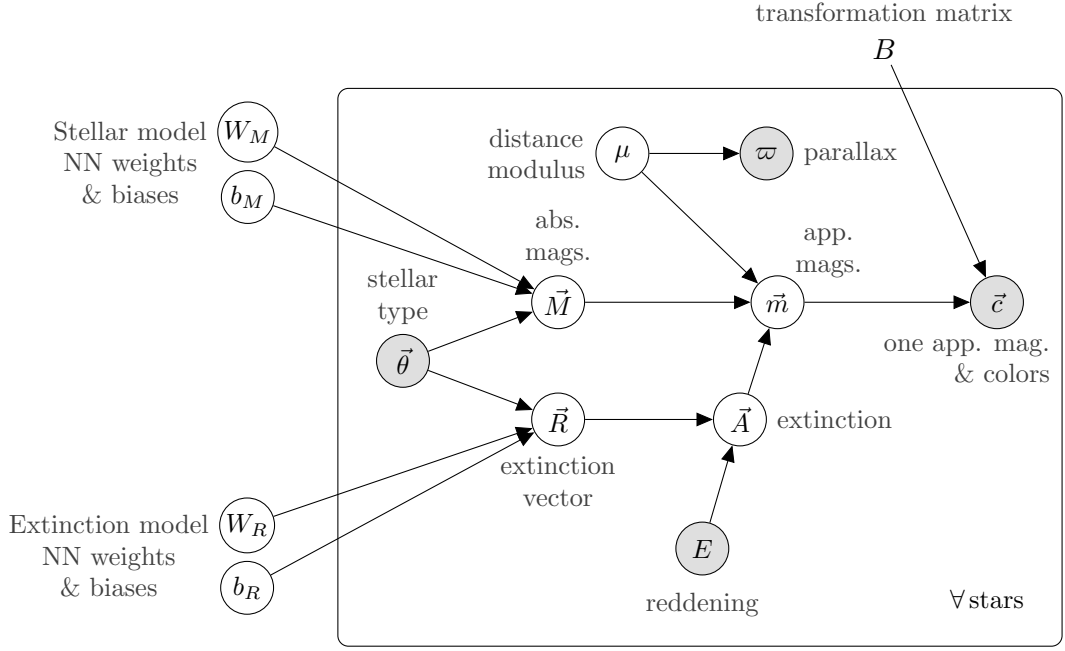
\begin{figure*}
    \begin{center}
        \begin{tikzpicture}[x=2.2cm, y=1.8cm]
            \node[obs,
                  label={[align=center]90:stellar\\type}]
                  (theta)
                  {$\vec{\theta}$} ;
            
            \node[latent, above right=of theta, yshift=-1.0cm,
                  label={[align=center]90:abs.\\mags.}]
                  (M)
                  {$\vec{M}$} ;
            \node[latent, right=of M,
                  label={[align=center,xshift=-0.15cm,yshift=-0.05cm]45:app.\\mags.}]
                  (m)
                  {$\vec{m}$} ;

            \node[latent, above left=of m, yshift=0.2cm,
                  label={[align=center]180:distance\\modulus}]
                  (mu)
                  {$\mu$} ;
            \node[obs, right=of mu, xshift=-1.0cm,
                  label={0:parallax}]
                  (varpi)
                  {$\varpi$} ;
            
            \node[obs, right=of m,
                  label={[xshift=-0.4cm,align=right]270:{one app. mag.\\ \& colors}}]
                  (c)
                  {$\vec{c}$} ;
            
            \node[latent, below right=of theta, yshift=1.0cm,
                  label={[align=center]270:{extinction\\ vector}}]
                  (R)
                  {$\vec{R}$} ;
            \node[latent, right=of R, xshift=-0.5cm,
                  label={[align=left]0:extinction}]
                  (A)
                  {$\vec{A}$} ;
            \node[obs, below=of A, xshift=-0.5cm, yshift=0.8cm,
                  label={[align=center,yshift=-0.1cm]270:reddening}]
                  (E)
                  {$E$} ;
            
            \node[latent, above left=of theta, yshift=0.4cm]
                  (bM)
                  {$b_M$} ;
            \node[latent, above=of bM, yshift=-1.7cm,
                  label={[yshift=-0.4cm,align=center]180:{Stellar model\\ NN weights\\ \& biases}}]
                  (WM)
                  {$W_M$} ;
            
            \node[latent, below left=of theta, yshift=-0.4cm,
                  label={[yshift=-0.4cm,align=center]180:{Extinction model\\ NN weights\\ \& biases}}]
                  (WR)
                  {$W_R$} ;
            \node[latent, below=of WR, yshift=1.7cm]
                  (bR)
                  {$b_R$} ;
            
            \node[const, above=of c, yshift=1.0cm, xshift=-1.5cm,
                  label={[yshift=1pt]90:transformation matrix}]
                  (B)
                  {$B$} ;

            \edge {theta,WM,bM} {M} ;
            \edge {theta,WR,bR} {R} ;
            \edge {R,E} {A} ;
            \edge {mu} {varpi} ;
            \edge {A,M,mu} {m} ;
            \edge {m,B} {c} ;

            \plate[inner sep=0.5cm]
              {star}
              {(theta) (M) (A) (mu) (R) (E) (c) (varpi)}
              {$\forall \, \mathrm{stars}$} ;
        \end{tikzpicture}
    \end{center}
    \caption{Directed factor graph \citep{Dietz2010} of our model of stellar magnitudes and colors. Shaded nodes represent ``observed'' quantities (i.e., the quantities with Gaussian estimates in our input catalog). Our model parameters are the neural network weights and biases that control the mapping from stellar type to absolute magnitudes, as well as the neural network weights and biases that control the dependence of the extinction vector on stellar type. The extinction vector controls the direction in magnitude-space that increasing dust density moves stars, and is only weakly dependent on stellar type, due to strong regularization of the weights $W_R$. The constant matrix $B$, given by Eq.~\eqref{eqn:transformation-matrix}, transforms the vector $\vec{m}$, containing apparent magnitudes, to the vector $\vec{c}$, containing one apparent magnitude and $\left(\mathrm{\# \ of\ bands}\right) - 1$ colors.}
    \label{fig:graphical-model}
\end{figure*}

Our goal is to learn a mapping $\vec{M} \args{\vec{\theta}\,}$ from spectroscopic features to absolute magnitude, as well as a vector $\vec{R}$ that describes how dust affects stellar magnitudes in each passband. We assume that extinction is given by $\vec{A} = E \, \vec{R}$, where $E$ is dependent on the amount of dust in front of the star. For an infinitesimally narrow passband, the vector $\vec{R}$ would not depend on spectroscopic features, and would be purely a property of the dust. For realistic passbands, however, $\vec{R}$ depends weakly on the stellar source spectrum (at the $\sim$10\% level for Gaia $G$ band, and less for other bands used in this work -- see Appendix \ref{app:nonlinear-extinction}). We will therefore model $\vec{R} \args{\vec{\theta}\,}$, with strong regularization to ensure that the dependence on the spectroscopic features is small. With these components, our forward model for observed stellar photometry is given by
\begin{align}
    \vec{m} &=
      \vec{M} \args{\vec{\theta}\,}
      + \mu
      + E \, \vec{R} \args{\vec{\theta}\,}
    \, ,
    \label{eqn:model-no-errors}
\end{align}
where $\mu$ is the distance modulus. The causal structure of our model is represented graphically in Fig.~\ref{fig:graphical-model}.

It is worthwhile to reflect for a moment on why it should be possible to learn such a model. We know from stellar astrophysics that most of the variation in stellar properties is determined by three parameters: initial mass, initial metallicity and age. There are higher-order parameters that affect stellar properties to a smaller extent (such as rotation and individual elemental abundances). Consequently, stars lie -- to rough approximation -- in a three-dimensional space. As initial mass and age are not easily measurable, they are not available for a large enough sample of stars to train a data-driven model of stellar photometry. It is, however, possible to measure a different set of parameters, describing the instantaneous present-day properties of the stellar atmosphere, from spectra: effective temperature, surface gravity and metallicity. While these spectroscopic features determine stellar colors, there is no guarantee that they will map to a unique luminosity, as the latter depends on stellar radius. Fortunately, the mapping from $\left( \teff, \logg, \feh \right)$ to radius turns out to be non-degenerate over most of the domain, so that \textit{absolute magnitudes} are nearly uniquely determined by these three features. In addition, the effect of dust extinction along the line-of-sight is to move stars along a single axis in multi-band magnitude space. Thus, in the absence of observational uncertainties, we would expect stars to lie along a 4-dimensional manifold in magnitude space. Though we take a data-driven approach in this work, the structure of our model -- with unextinguished stellar photometry lying along a three-dimensional manifold in magnitude space, and extinction moving stars along a nearly type-independent axis in magnitude space -- is informed by the underlying physics.

We will learn both the mapping $\vec{M} ( \vec{\theta}\, )$ and extinction vector $\vec{R} ( \vec{\theta}\, )$ from a training dataset of stars with spectroscopically determined type, $\hat{\theta}$ (for example, $\teff$, $\logg$, and $\feh$, though our method does not depend on the precise choice of parameterization of stellar type), observed apparent magnitudes $\hat{m}$, measured reddening $\hat{E}$, and for a subset of the stars, observed parallax $\hat{\varpi}$. We defined our model above in Eq.~\ref{eqn:model-no-errors}, assuming that $\vec{\theta}$, $\mu$ and $E$ were exactly known. As this is not true for realistic observations, we Taylor expand our model to first order in the observational errors:
\begin{align}
    \hat{m} =&\,
        \vec{M} ( \hat{\theta} )
      + \mu \arg{\hat{\varpi}}
      + \hat{E} \, \vec{R} ( \hat{\theta} )
    \notag \\
    &
      \ \ - \ 
        \left(
          \delta \vec{\theta} \cdot
          \nabla_{\vec{\theta}}
        \right) \!\!
        \left(
          \vec{M} + \hat{E} \, \vec{R}
        \right)
        \! \bigg|_{\vec{\theta} = \hat{\theta}}
      \!\!\!\!\!
      - \,\,
        \delta \varpi \left.
          \frac{\partial \mu}{\partial \varpi}
        \right|_{\varpi=\hat{\varpi}}
      \!\!\!\!\!\!\!
      - \,\, \delta E \, \vec{R}
      + \delta \vec{m}
    \, ,
    \label{eqn:model-with-errors}
\end{align}
where hatted quantities are noisy measurements of the corresponding un-hatted quantities, and $\delta \vec{m}$, $\delta \vec{\theta}$, $\delta \varpi$ and $\delta E$ are the errors ($\mathrm{measurement} - \mathrm{truth}$) in the corresponding measured quantities. This first-order expansion is valid when $\delta \vec{\theta}$ is small and $\delta \varpi \ll \varpi$.

\begin{figure}[ht!]
  \epsscale{0.8}
  \plotone{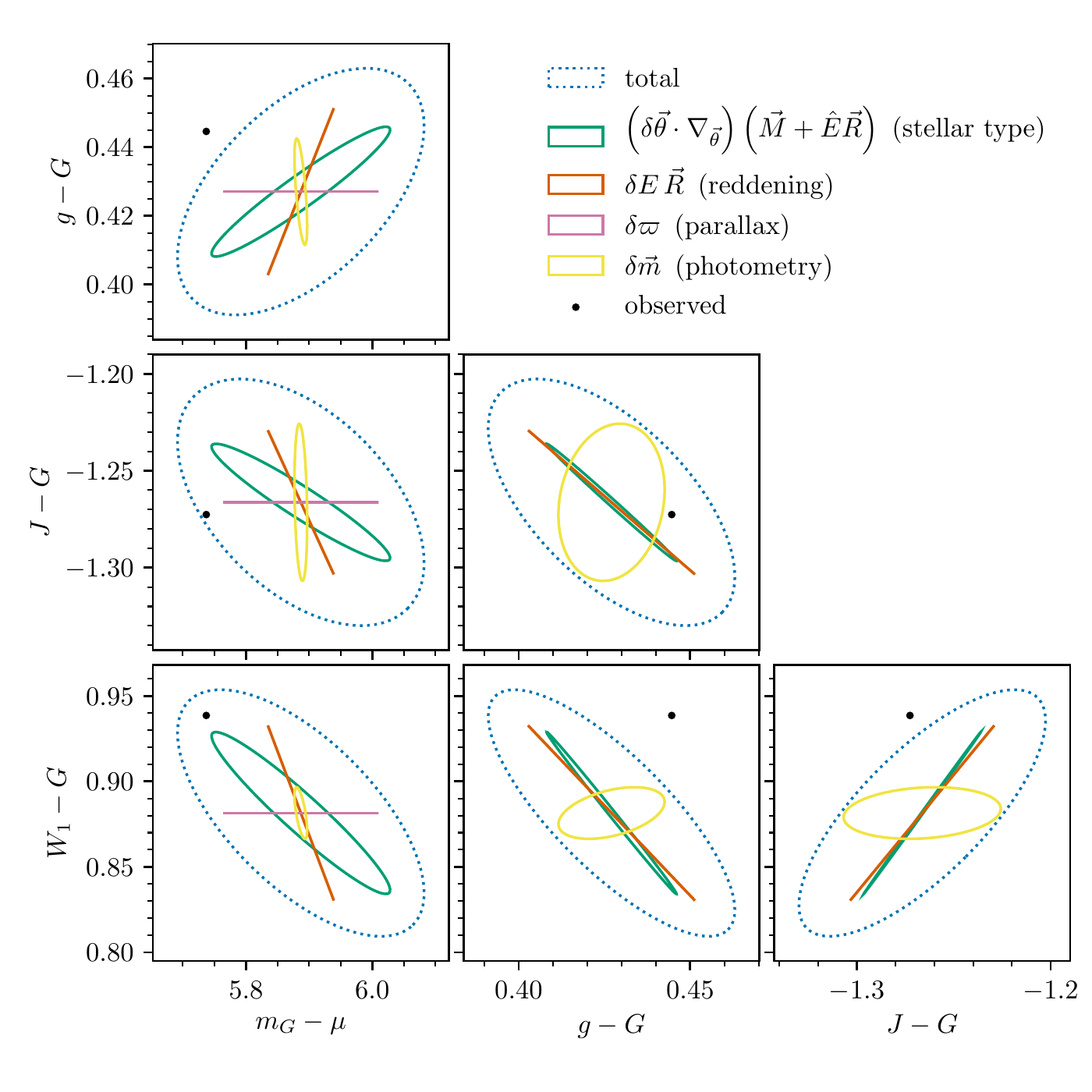}
  \caption{Components of the covariance matrix for an example star, in a subset of the photometric passbands used in this work. Each ellipse corresponds to the $1\sigma$ region for a particular source of error, such as uncertainty in the stellar type or reddening. The ellipses are centered on the model prediction for the star's photometry, while the observed photometry is marked by a black dot. \label{fig:covariance-components}}
\end{figure}

Assume that the observational errors, $\delta E$, $\delta \varpi$, $\delta \vec{\theta}$ and $\delta \vec{m}$, have Gaussian distributions and are independent of one another, with (co)variances of $\sigma_E^2$, $\sigma_{\varpi}^2$, $C_{\theta}$ and $\vec{\sigma}_m^2$, respectively. In the first-order Taylor approximation, the observed apparent magnitudes $\hat{m}$ are then also Gaussian-distributed, with covariance given by
\begin{align}
  C_{m ,\, ij} &=
      \sum_{k\ell}
        \frac{\partial (M_i \!+\! \hat{E} R_i)}{\partial \theta_k}
        \frac{\partial (M_j \!+\! \hat{E} R_j)}{\partial \theta_{\ell}}
        \, C_{\theta ,\, k \ell}
    + R_i R_j \sigma_E^2
    + \left| \frac{5}{\ln 10} \frac{\sigma_{\varpi}}{\hat{\varpi}} \right|^2
    + \delta_{ij} \sigma_{m , i}^2 \, ,
    \label{eqn:covariance-matrix}
\end{align}
where $i$ and $j$ refer to passbands $i$ and $j$, and $\delta_{ij}$ is the Kronecker delta. Using this covariance matrix, we can calculate a Gaussian likelihood for any observed star in our training dataset. Fig~\ref{fig:covariance-components} shows the contributions of each term to the covariance matrix for an example star. Note that this covariance matrix depends on the gradients of our model for $\vec{M} ( \vec{\theta} \, )$ and $\vec{R} ( \vec{\theta} \, )$, as well as on the value of $\vec{R} ( \vec{\theta} \, )$. As we adjust the mappings $\vec{M} ( \vec{\theta} \, )$ and $\vec{R} ( \vec{\theta} \, )$, the covariance matrix of $\hat{m}$ must also change.

This method works well if $\sigma_{\varpi} \ll \varpi$. As parallax uncertainty grows, the uncertainty in distance modulus $\mu$ becomes increasingly non-Gaussian, and we can no longer assume the uncertainty in apparent magnitude to be Gaussian. However, this non-Gaussianity does not propagate into stellar \textit{colors}, as they do not depend on distance modulus. Instead of modeling $\vec{m}$, we model one apparent magnitude and $\left(\mathrm{\#\ of\ bands} - 1\right)$ colors. We define $\vec{c} \equiv B \, \vec{m}$, where
\begin{align}
  B &=
    \begin{pmatrix}
       1 & 0 & 0 & 0 & \hdots \\
      -1 & 1 & 0 & 0 & \hdots \\
      -1 & 0 & 1 & 0 & \hdots \\
      -1 & 0 & 0 & 1 &        \\
      \vdots & \vdots & \vdots & & \ddots
    \end{pmatrix} \, .
  \label{eqn:transformation-matrix}
\end{align}
This makes the first entry in $\vec{c}$ the apparent magnitude of the first passband, while the rest of the entries are colors: $c_i = m_i - m_0$. The covariance matrix describing the uncertainty in $\vec{c}$ is given by
\begin{align}
  C_c = B \, C_m \, B^T \, .
  \label{eqn:covariance-transformed}
\end{align}
After this transformation, the non-Gaussianity stemming from the parallax-to-distance-modulus conversion is confined to the first element of $\vec{c}$, $c_0$, representing apparent magnitude. For stars with large uncertainties in $\varpi$, we only model stellar colors, ignoring $c_0$. Thus, we learn color information from all stars, but only learn about the \textit{absolute} magnitudes of stars from those with well-measured parallaxes.

This now leaves us to find the functions $\vec{M} \args{\vec{\theta}\,}$ and $\vec{R} \args{\vec{\theta}\,}$ that maximize the likelihood of our stellar data:
\begin{align}
  \pcond{
    \left\{ \hat{c} \right\}
  }{
    \left\{ \hat{\theta} ,\, \hat{\mu} ,\, \hat{E} \right\}
  }
  &=
  \prod_{\mathrm{star}\,i}
  \pcond{
    \hat{c}_i
  }{
    \hat{\theta}_i ,\, \hat{\mu}_i ,\, \hat{E}_i
  }
  =
  \prod_{\mathrm{star}\,i}
  \mathcal{N} \left(
    \hat{c}_i
  \mid
    B \left< \hat{m} \right>_i ,\,
    C_{c,i}
  \right)
  \, ,
  \label{eqn:likelihood}
\end{align}
where $\left< \hat{m} \right> = \vec{M} \args{\hat{\theta}} + \mu \arg{\hat{\varpi}} + \hat{E} \, \vec{R} \args{\hat{\theta}}$. Many classes of fitting functions could be used to model $\vec{M} ( \vec{\theta} \, )$, such as polynomials or Gaussian processes. Here, we employ a shallow feed-forward neural network. In this context, the neural network should simply be thought of as a highly flexible, regularized fitting function. Our model for $\vec{R} \args{\vec{\theta}\,}$ is the exponential of a linear function of the spectroscopic features, with strong regularization of the linear weights. In the limit that the weights go to zero, $\vec{R}$ becomes a constant vector, independent of the spectroscopic features. We apply an exponential activation function to the layer, which forces the entries of $\vec{R}$ to be positive. Our stellar model is thus a highly flexible fitting function, while our extinction model is highly regularized and expected to be approximately represented by a constant vector. We use the negative logarithm of the likelihood (Eq.~\ref{eqn:likelihood}) as the loss function, with additional regularization terms described below.

As noted earlier, the covariance matrix (Eq.~\ref{eqn:covariance-matrix}) requires knowledge of the gradients of the model functions $\vec{M} \args{\vec{\theta}\,}$ and $\vec{R} \args{\vec{\theta}\,}$. With a trained neural network representing the model, these gradients are easy to calculate for any $\hat{\theta}$. However, when we begin training the model, we do not know the values of these gradients. We therefore iterate between training the model while holding the covariance matrices $\left\{ C_m \right\}$ fixed, and updating the covariance matrices $ \left\{ C_m \right\}$ while holding the model fixed. When calculating the covariance matrices before the initial iteration, we assume that the gradients are zero and that $\vec{R} = 0$, meaning that only uncertainties in observed photometry and parallax contribute to $C_m$. In iteration $n$, we use the covariance matrices $ \left\{ C_m \right\}$ determined using the model resulting from iteration $n-1$.

After each iteration, we additionally update our estimates of the reddening of each star. Given a model $\vec{M} \args{\vec{\theta}\,}$ and $\vec{R} \args{\vec{\theta}\,}$, we can calculate the predicted zero-extinction magnitudes of a star, $\vec{m}_{\mathrm{pred}}$, by setting $\hat{E} = 0$ in Eq.~\eqref{eqn:model-with-errors}. The covariance matrix of this prediction, $C_{\mathrm{pred}}$, is given by Eq.~\eqref{eqn:covariance-matrix}, with $\hat{E} = \sigma_E = 0$. We can then infer the updated stellar reddening $E^{\prime}$ using the difference between the observed and predicted zero-extinction magnitudes, taking into account the covariance matrix $C_{\mathrm{pred}}$. If we put a Gaussian prior on $E^{\prime}$, with mean $\hat{E}$ and variance $\sigma_{0}^2$, then the resulting posterior on $E^{\prime}$ is Gaussian, with mean and variance given by
\begin{align}
    \left< E^{\prime} \right> &= \frac{
        \frac{E_0}{\sigma_0^2}
      + \vec{R}^{\,T} C_{\mathrm{pred}}^{-1} \left(
          \hat{m} - \vec{m}_{\mathrm{pred}}
        \right)
    }{
        \vec{R}^{\,T} C_{\mathrm{pred}}^{-1} \vec{R}
      + \frac{1}{\sigma_0^2}
    }
    \label{eqn:E-estimate}
    \, , \\
    \sigma_{E^{\prime}}^2 &=
      \left(
          \vec{R}^{\,T} C_{\mathrm{pred}}^{-1} \vec{R}
        + \frac{1}{\sigma_0^2}
      \right)^{\! -1}
    \, .
    \label{eqn:sigma_E-estimate}
\end{align}
We use the initial estimate of reddening to set the mean and variance of our prior. Keeping the mean of the prior constant across iterations is important in order to avoid a degeneracy in the model between individual stellar reddenings and the length of the vector $\vec{R}$. If we were to update the prior after each iteration, the reddenings of all the stars in the training dataset could drift either up or down, compensated exactly by an inverse scaling of $\vec{R}$. We clip the inferred reddenings, such that $\left< E^{\prime} \right> \geq 0\,\mathrm{mag}$. We additionally impose a floor of ${\sigma_{E^{\prime}}^2 \geq \left( 0.02 \,\mathrm{mag} \right)^2 + \left( 0.1 \, E^{\prime} \right)^2}$ on the inferred reddening variances.

\subsection{Neural Network Structure}
\label{sec:neural-network}

\begin{figure*}[ht!]
  \epsscale{1.1}
  \plotone{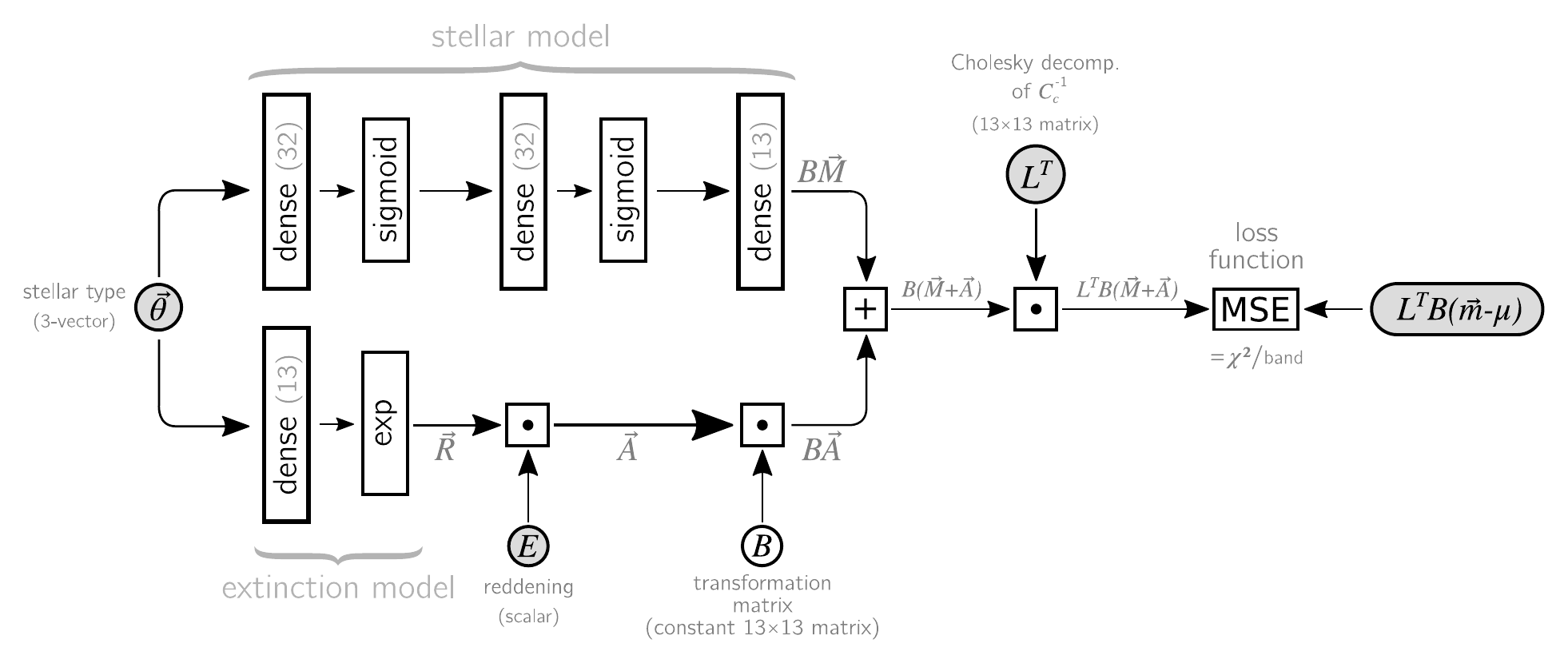}
  \caption{Our neural network structure. The shaded elements $\vec{\theta}$, $E$ and $L^T$ are inputs. The network outputs its prediction of $L^T B \left( \vec{M} \! + \! \vec{A} \right)$. This is compared to the observed quantity $L^T B \left( \vec{m} \! - \! \mu \right)$ (shaded, on the right of the figure), using a mean square error (MSE) loss function. This is equivalent to calculating the $\chi^2 / \mathrm{band}$ of the predicted magnitudes and colors, as shown in Eqs.~\eqref{eqn:chisq-begin}--\eqref{eqn:chisq-end}. Light gray text indicates the physical interpretation of the outputs of individual layers and sectors of the neural network. The dense layers in the top half of the diagram, representing the stellar model, have L2 weight regularization of $10^{-4}$. The dense layer in the bottom half of the diagram outputs the vector $\vec{R}$, and has an exponential activation function, in order to ensure that $\vec{R}$ only has positive entries. The weights of this dense layer have L1 weight regularization of $10^{-2}$, so that the model \textit{a priori} prefers to $\vec{R}$ to be independent of the spectroscopic features $\vec{\theta}$. $\vec{R}$ is multiplied by the input $E$ to obtain extinction $\vec{A}$. The size of each dense layer is indicated in parentheses. The final dense layer in the stellar model and the dense layer representing the extinction vector have size 13, equal to the number of photometric passbands in this work. \label{fig:neural-network}}
\end{figure*}

The structure of our neural network is illustrated in Fig.~\ref{fig:neural-network}. The vector $\vec{\theta}$ contains the spectroscopic features, with $\teff$, $\logg$ and $\feh$ each shifted and scaled so that the entire input dataset has a median of zero and a standard deviation of unity. The $\vec{M} \args{\vec{\theta}\,}$ sector of the neural network takes in the vector $\hat{\theta}$, passing it through two hidden layers to obtain the absolute magnitude in the first passband and the colors relative to the first passband: $B \, \vec{M}$. The $\vec{R} \args{\vec{\theta}\,}$ sector of the neural network feeds $\hat{\theta}$ through a single dense layer with exponential activation (in order to ensure positivity), to obtain the vector $\vec{R}$. This vector is multiplied by the input $\hat{E}$ to obtain extinction, and then transformed to a mixture of magnitude- and color-space using the matrix $B$. The network adds the output of the two sectors to obtain $B \big( \vec{M} + \hat{E} \vec{R} \, \big)$, which by Eq.~\eqref{eqn:model-with-errors} should be equal to $B \left[ \hat{m} - \mu \arg{\hat{\varpi}} \right]$, up to an error term with covariance $C_c$, given by Eq.~\eqref{eqn:covariance-transformed}.

Let $L$ be the Cholesky decomposition of $C_c^{-1}$:
\begin{align}
    L \, L^T = C_c^{-1}.
\end{align}
We operate with $L^T$ on the output of the network, obtaining $L^T B ( \vec{M} + \hat{E} \vec{R} )$. We compare this output with our expected output, $L^T B \left[ \hat{m} - \mu \arg{\hat{\varpi}} \right]$, using the mean square error loss function. This is equal to the $\chi^2$ error corresponding to the stellar likelihood, divided by the number of passbands:
\begin{align}
  \label{eqn:chisq-begin}
  \chi^2 &=
    \left|
      L^T B \left( \vec{M} + \hat{E} \vec{R} \right)
      -
      L^T B \left[ \hat{m} - \mu \arg{\hat{\varpi}} \right]
    \right|^2
  \\
  &=
    \left|
      L^T
    \right.
      \underbrace{
        B
        \left[
          \vec{M} + \hat{E} \vec{R} + \mu \arg{\hat{\varpi}} - \hat{m}
        \right]
      }_{
        \equiv \Delta \vec{c}
      }
    \left.
      \vphantom{L^T}
    \right|^2
  \\
  &= \Delta \vec{c}^{\, T} L L^T \Delta \vec{c}
  \\
  &= \Delta \vec{c}^{\, T} C_c^{-1} \Delta \vec{c}
  \, .
  \label{eqn:chisq-end}
\end{align}
Minimizing the mean square error loss between the output of the network and the expected output is thus equivalent to maximizing the stellar likelihood. During training, the network varies both the stellar model and the extinction vector in order to minimize the sum of $\chi^2$ over all the stars (equivalent to maximizing the product of all the stellar likelihoods).

The loss function contains regularizing terms in order to prevent the model from over-fitting the data. In the dense layers that make up the stellar model (see Fig.~\ref{fig:neural-network}), we apply an L2 weight regularization of $10^{-4}$ to each dense layer. In the dense layer of the extinction model, we instead apply an L1 weight regularization (of $10^{-2}$), which penalizes even small variations in the reddening vector $\vec{R}$, encouraging it to be independent of the spectroscopic features. Specifically, the per-star loss function is
\begin{align}
    \mathcal{L}
    &=
    \frac{\chi^2}{n_{\mathrm{bands}}}
    + \lambda_s \! \sum_{w \in W_s} \! w^2
    + \lambda_{r} \! \sum_{w \in W_r} \left| w \right|
    \, ,
    \label{eqn:loss-with-regularization}
\end{align}
where $n_{\mathrm{bands}} = 13$ is the number of photometric passbands, $W_s$ is the set of all weights in the stellar model, $W_r$ is the set of all weights in the extinction model, $\lambda_s = 10^{-4}$ and $\lambda_{r} = 10^{-2}$. During training, we minimize the value of this loss function, averaged over the stars in the training dataset.

\section{Data}
\label{sec:data}

\subsection{Spectroscopic Features}
\label{sec:spectroscopic-features}

We obtain stellar parameters from three spectroscopic surveys: the Data-Driven Payne reduction \citep[``DDPayne'';][]{Maosheng2019DDPayne} of LAMOST DR5, APOGEE \citep[SDSS-IV DR16,][]{Ahumada2019APOGEEDR16} and GALAH \citep[DR2,][]{Buder2018GALAHDR2}.

The Large Sky Area Multi-Object Fiber Spectroscopic Telescope \citep[``LAMOST'';][]{Cui2012LAMOST,Zhao2012LAMOST} is a high-\'{e}tendue telescope with a low-resolution ($R \sim 1,800)$ spectrograph, which has surveyed millions of bright Milky Way stars. \citet{Maosheng2019DDPayne} develops a data-driven method, the ``Data-Driven Payne,'' that builds upon ``The Payne'' \citep{Ting2019Payne}, and applies this method to spectra of over six million stars in LAMOST Data Release 5 (DR5), obtaining $\teff$, $\logg$, $v_{\mathrm{mic}}$ and 16 elemental abundances. We use a diagonal covariance matrix $C_{\theta}$ for LAMOST sources, using the variances reported by the Data-Driven Payne. We require that $\mathrm{SNR} > 20$ in $u$, $g$, $r$, $i$ or $z$ band, and that $\mathtt{qflag\_chi2} == `\mathtt{good}\textrm'$ and $\mathtt{flag\_singlestar} == `\mathtt{YES}\textrm'$.

The Apache Point Observatory Galactic Evolution Experiment \citep[``APOGEE'';][]{Majewski2017APOGEE}, part of SDSS-III, is a survey conducted on the Sloan 2.5~m Telescope, using a $R \sim 22,500$ near-infrared spectrograph. As part of SDSS-IV, APOGEE-2 \citep{Majewski2017APOGEE} continues to survey the North using the original spectrograph, and has begun observations of the Southern sky using a duplicate spectrograph installed on the 2.5~m du Pont telescope at Las Campanas Observatory. SDSS Data Release 16 \citep[DR16;][]{Ahumada2019APOGEEDR16} contains APOGEE and APOGEE-2 spectra of approximately 430,000 stars. The APOGEE Stellar Parameter and Chemical Abundances Pipeline \citep[``ASPCAP'';][]{GarciaPerez2016ASPCAP} estimates stellar parameters and abundances. ASPCAP reports a metallicity $\left[ \mathrm{M} / \mathrm{H} \right]$, which we treat as $\feh$. ASPCAP reports two sets of uncertainties: a covariance matrix based on a regression to synthetic stellar spectral models, and empirically calibrated uncertainties. The latter are generally larger than the former, but do not contain off-diagonal covariance elements. For APOGEE sources, we set the off-diagonal elements of the covariance matrix $C_{\theta}$ using the uncalibrated covariance matrix, and set the diagonal elements to whichever is larger: the calibrated or uncalibrated variances. We cut out APOGEE sources for which either the \texttt{STAR\_WARN} or \texttt{STAR\_BAD} bit is set.

The Galactic Archaeology with HERMES \citep[``GALAH'';][]{DeSilva2015GALAH,Martell2017GALAH} survey uses the HERMES spectrograph on the Anglo-Australian Telescope to take $R \sim 28,000$ spectra in four optical bands, primarily targeting FGK stars in the Galactic disk. GALAH Data Release 2 (DR2) contains spectra of more than 340,000 stars, and delivers stellar parameters including $\teff$, $\feh$, $\logg$, $v \sin i$, $v_{\mathrm{mic}}$, $A_{K_s}$ and multiple elemental abundances \citep{Buder2018GALAHDR2}. As GALAH DR2 does not report covariances between the spectroscopic features, we set the covariance matrix $C_{\theta}$ to be diagonal for GALAH sources. We cut out GALAH sources with low signal-to-noise ratios (SNR), requiring $\mathrm{SNR} > 20$ in at least one of GALAH's four channels. We also require that $\mathtt{flag\_cannon} == 0$.

Using these three surveys allows us to cover a wide range of stellar types in the range $4000\,\mathrm{K} \lesssim \teff \lesssim 7500\,\mathrm{K}$. One difficulty that using multiple spectroscopic surveys creates, however, is that the estimates of stellar type may vary systematically between the surveys, due to both differences in the instruments themselves and differences in the data reduction pipelines. Our aim is to learn a mapping from stellar type, $\vec{\theta}$, to absolute magnitudes, $\vec{M}$. If one learns the mapping $\vec{\theta}_a \rightarrow \vec{M}$ with data from spectroscopic survey $a$, but then inputs $\vec{\theta}_b$ measured by another spectroscopic survey with systematic differences in its stellar parameter estimates, the resulting predictions of $\vec{M}$ will be systematically off. We require that the different spectroscopic surveys provide \textit{consistent} stellar types, but not that they provide the \textit{correct} stellar types. That is, for spectroscopic surveys $a$ and $b$ measuring the same star, we require that $\vec{\theta}_a \approx \vec{\theta}_b$. We are using stellar spectroscopic features as a coordinate system that spans the range of possible stellar types, and we require that each survey map the same star to the same location in this coordinate system. Even if the input $\hat{\theta}$ vectors do not correspond exactly to true $\teff$, $\logg$ and $\feh$, we will still learn the mapping from $\hat{\theta}$ to photometry.

We therefore add zero-point corrections to the APOGEE and LAMOST spectroscopic features, so that their estimates of $\teff$, $\logg$ and $\feh$ match those of GALAH, on average. We separately match APOGEE and GALAH to LAMOST, and then compute the inverse-variance-weighted average offset in $\teff$, $\logg$ and $\feh$ between APOGEE and LAMOST and between GALAH and LAMOST. We then use the resulting offsets to bring APOGEE and LAMOST onto the same scale as GALAH. This is the lowest-order correction that one can make. More complicated corrections -- including $\vec{\theta}$-dependent corrections or recalibration of reported uncertainties -- are possible, and could even be included as hyperparameters in our model. However, these more complicated corrections are beyond the scope of this paper. \textit{What we are ultimately learning is therefore a mapping from the stellar spectroscopic features -- with calibration systematics tied to GALAH DR2 -- to stellar photometry}.

The offsets that we add into LAMOST spectroscopic features are $\left( \Delta\teff ,\, \Delta\logg ,\, \Delta\feh \right) = \left( -3.6\,\mathrm{K} ,\, -0.014\,\mathrm{dex} ,\, 0.068\,\mathrm{dex} \right)$, while the offsets that we add into APOGEE are $\left( \Delta\teff ,\, \Delta\logg ,\, \Delta\feh \right) = \left( -26.6\,\mathrm{K} ,\, -0.026\,\mathrm{dex} ,\, 0.018\,\mathrm{dex} \right)$.

In order to avoid unrealistically low uncertainties on the spectroscopic features, we add $\left( 10\,\mathrm{K} \right)^2$, $\left( 0.05\,\mathrm{dex} \right)^2$ and $\left( 0.03\,\mathrm{dex} \right)^2$ to the variances in $\teff$, $\logg$ and $\feh$, respectively. Finally, we cut sources with standard deviation greater than 200~K in $\teff$, or 0.5~dex in $\logg$ or $\feh$.

\subsection{Photometry and Parallaxes}
\label{sec:photometry-parallax}

\begin{figure*}[ht!]
\plotone{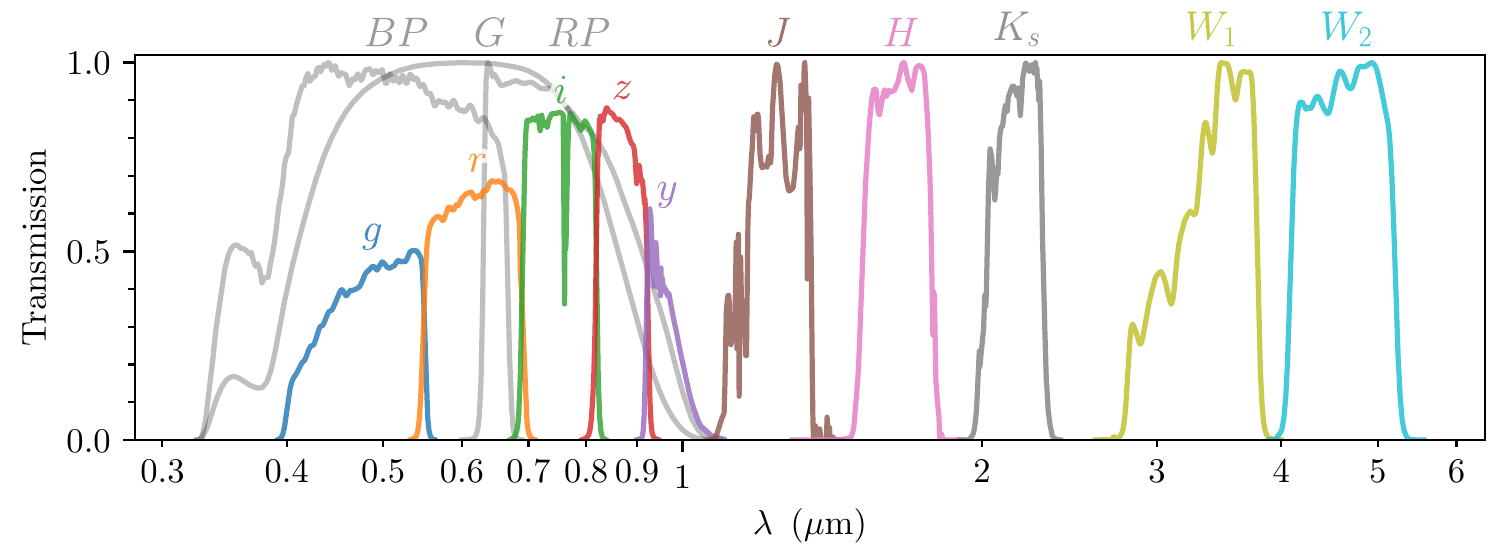}
  \caption{Transmission curves of the photometric passbands used in this work. The peak transmission of Gaia, 2MASS and WISE passbands are normalized to unity. We plot the Gaia transmission curves calculated by \citet{ApellanizWeiler2018}, the Pan-STARRS~1 transmission measured by \citet{PS1PhotometricSystem}, the 2MASS transmission measured by \citet{Cohen2003}, and WISE transmission determined by \citet{Wright2010}. All transmission curves were obtained through the Spanish Virtual Observatory \citep{SVO1,SVO2}. \label{fig:filter-curves}}
\end{figure*}

Our method can be applied to any arbitrary set of photometric passbands. We demonstrate our method using 13 photometric passbands, from Gaia ($G$, $BP$, $RP$), Pan-STARRS~1 ($g_{\mathrm{P1}}$, $r_{\mathrm{P1}}$, $i_{\mathrm{P1}}$, $z_{\mathrm{P1}}$, $y_{\mathrm{P1}}$), 2MASS ($J$, $H$, $K_s$) and unWISE ($W_1$, $W_2$), spanning wavelengths of $\sim 330\,\mathrm{nm}$ to $\sim 5000\,\mathrm{nm}$. Fig.~\ref{fig:filter-curves} shows how these passbands cover the observed wavelength range.

Gaia is a space telescope that is conducting a large-scale astrometric stellar survey, measuring parallaxes and proper motions of billions of stars to unprecedented precision \citep{GaiaCollaboration2016Mission}. In addition, Gaia provides photometry in a wide band, $G$, and spectrophotometry in two bands that essentially cover the two halves of the $G$ passband, $BP$ and $RP$ \citep{Liu2012Spectrophotometry}. Gaia Data Release 2 (DR2) contains five-parameter astrometric determinations for 1.3 billion sources \citep{Lindegren2018DR2Astrometry,GaiaCollaboration2018DR2Contents}. We make use of all three photometric passbands from Gaia, as well as Gaia's parallax measurement.

We choose Gaia $G$ as the first passband in our vector $\vec{c}$, meaning that $c_0$ represents $G$-band apparent magnitude, while the remaining entries in $\vec{c}$ represent colors with respect to $G$. Because we learn absolute magnitudes in the first passband, we would like stars with well-measured parallaxes to also be measured in this passband, making $G$ the natural choice. We additionally learn colors relative to the first passband, restricting our sample to stars that are detected in this passband. The range of apparent magnitudes probed by Gaia $G$ spans the range of apparent magnitudes of stars targeted by LAMOST, APOGEE and GALAH, so that stars observed by these spectroscopic surveys are likely to have Gaia $G$-band magnitudes. The ordering of the remaining passbands in $\vec{c}$ makes no difference to our method.

We make the following quality cuts on Gaia sources, as recommended by \citet{Arenou2018CatalogValidation}:
\begin{enumerate}
    \item $\quad \mathtt{visibility\_periods\_used} > 8$
    \item $\quad \frac{\mathtt{astrometric\_chi2\_al}}{\mathtt{astrometric\_n\_good\_obs\_al} - 5} < 1.44 \, \max \left\{ 1, \ \exp \left[ -0.4 \left( m_{G} - 19.5 \right) \right] \right\}$
\end{enumerate}
where $m_G$ is the Gaia $G$-band magnitude. The first cut removes sources with an insufficient number of Gaia observations to reliably constrain the astrometric model, while the second cut removes sources which are poorly fit by the astrometric model. We additionally require three or more observations in $G$ band: $\mathtt{g\_n\_obs} > 2$.

The Panoramic Survey Telescope and Rapid Response System 1 (Pan-STARRS~1, hereafter abbreviated as ``PS1'') is a 1.8~m optical and near-infrared telescope located on Mount Haleakala, Hawaii \citep{Chambers2016PS1}, equipped with the Gigapixel Camera \#1, consisting of an array of 60 CCD detectors, each 4800 pixels on a side \citep{Tonry2006-GPC1,Onaka2008-GPC1,Chambers2016PS1,Stubbs2010-PS1-lasercal}. From 2010 to 2014, PS1 conducted a multi-epoch, five-band ($g_{\mathrm{P1}}$, $r_{\mathrm{P1}}$, $i_{\mathrm{P1}}$, $z_{\mathrm{P1}}$, $y_{\mathrm{P1}}$) survey of the sky north of declination $\delta = -30^{\circ}$ \citep{Chambers2016PS1}. We use the mean of the photometric single-epoch PS1 photometry used to create Data Release 1 \citep{Flewelling2016-PS1-database}, excluding PS1 photometry flagged as unphotometric, saturated or containing bad pixels.

The Two Micron All Sky Survey (2MASS) is an all-sky near-infrared ($J$, $H$ and $K_{s}$) survey \citep{Skrutskie2006}, which made use of two 1.3~m telescopes, at Mount Hopkins, Arizona and Cerro Tololo, Chile. The focal plane of each telescope was equipped with three $256 \times 256$ pixel arrays, with a $2^{\prime}$ pixel scale. The entire sky was covered six times with dual 51-millisecond and 1.3-second exposures, in order to cover a wide range in apparent magnitudes. The survey achieved a 10-$\sigma$ point-source depth of $\sim$15.8, $\sim$15.1 and $\sim$14.3~mag (Vega) in $J$, $H$ and $K_{s}$, respectively. 2MASS photometry was calibrated to 0.02~mag accuracy, with per-source photometric uncertainties for bright sources below 0.03~mag \citep{Skrutskie2006}. We make use of the 2MASS ``high-reliability catalog''\footnote{See the \textit{Explanatory Supplement to the 2MASS All Sky Data Release}: \url{https://old.ipac.caltech.edu/2mass/releases/allsky/doc/sec1\_6b.html\#composite}}. When 2MASS reports that a source is extended, we remove it entirely from our training catalog.

The Wide-field Infrared Survey Explorer \citep[``WISE'';][]{Wright2010} is a 40~cm-aperture space telescope which in 2010 surveyed the full sky in four passbands, centered on 3.4, 4.6, 12 and 22~$\mathrm{\mu}m$ (and named $W_1$, $W_2$, $W_3$ and $W_4$, respectively), until the depletion of its cryogenic coolant. The spacecraft conducted several months of additional observations in $W_1$ and $W_2$, which do not require cryogenic cooling, in order to search for near-earth objects \citep[``NEOWISE'';][]{Mainzer2011}, and was subsequently put into hibernation. The spacecraft was reactivated in 2013 in order to continue the NEOWISE mission \citep{Mainzer2014}. Since then, WISE has gathered extensive photometry in the $W_1$ and $W_2$ bands across much of the sky. The ``unWISE'' Catalog \citep{Schlafly2019} is based on a reprocessing of all publicly available WISE $W_1$ and $W_2$ photometry. This reprocessing is both deeper than previous WISE catalogs, due to the co-addition of a larger number of observations, and performs better in the Galactic plane, due to its use of a photometric pipeline optimized for crowded fields \citep{Schlafly2018}. We use unWISE photometry for which $\mathtt{flags} == 0$ and $\mathtt{fracflux} > 0.5$, and reject photometric bands for which $\chi^2 / \mathrm{d.o.f.} > 5$. We convert unWISE magnitudes from Vega to the AB system by adding 2.699~mag and 3.339~mag to the reported $W_1$ and $W_2$ magnitudes, respectively \citep{Schlafly2019}.

In order to remove unrealistically small photometric uncertainties, we transform the reported uncertainties in the passbands from all surveys by adding 0.02~mag in quadrature. We handle unobserved passbands (and those that fail our quality cuts) by setting the photometric uncertainty to infinity. We additionally treat any passband with uncertainty greater than 0.2~mag as if it were unobserved. In order to limit the non-Gaussian error effects described in \S\ref{sec:method} and Appendix \ref{app:noisy-parallax}, which occur for sources with weakly constrained parallaxes, we disregard parallax measurements for which $\hat{\varpi} / \sigma_{\varpi} < 5$. For these sources, we set $\sigma_{\varpi}$ to infinity. This renders the uncertainty in absolute magnitude infinite, but does not affect the uncertainties in the colors.

\subsection{Reddenings}
\label{sec:reddenings}
  
As discussed in \S\ref{sec:method}, our method requires a Gaussian estimate of the stellar reddening for each source: ($\hat{E}$,~$\sigma_E$). We use a combination of reddening measurements from \citet[][``SFD'']{SFD} and \citet[][``Bayestar19'']{Green2019} to set $\hat{E}$ and $\sigma_E$ for each star. By construction, these two dust maps report reddening in equivalent units, so that they can be used interchangeably.

For stars that are more than 400~pc from the midplane of the Galaxy, we set mean reddening to the SFD estimate, and the standard deviation to 10\% of the SFD reddening. For the purpose of determining whether a given star is more than 400~pc from the midplane, we use the $5\sigma$ upper bound on parallax to determine the distance to the star. We set the reddenings of 40\% of the stars in our input catalog in this manner.
 
For stars that are closer to the midplane, it is important to take into account the three-dimensional distribution of dust. For stars that lie in the Bayestar19 footprint (declination greater than $-30^{\circ}$), have high signal-to-noise Gaia parallax measurements ($\hat{\varpi}/\sigma_{\varpi} > 5$) and reside less than 400~pc from the midplane, we therefore use the Bayestar19 reddening estimate. For each star, we sample parallaxes from the Gaia likelihood, naively transform those parallaxes to distances, and draw a sample from the Bayestar19 reddening estimate at each sampled distance. We use the mean and standard deviation of the resulting Bayestar19 reddening samples to set $\hat{E}$ and $\sigma_E$. We add an additional reddening uncertainty, equal to 10\% of the Bayestar19 reddening estimate, in quadrature. We set the reddenings of 51\% of the stars in our input catalog using Bayestar19.

For all other stars, we set $E = 0\,\mathrm{mag}$, and set $\sigma_E$ to the SFD reddening at the location of the given star. This category includes stars in the midplane of the Galaxy but outside the footprint of the Bayestar19 map (i.e., declination less than $-30^{\circ}$). We set the reddenings of 9\% of the stars in our input catalog in this manner.

In order to avoid unrealistically low reddening uncertainties, we add an additional 0.02~mag in quadrature to the $\sigma_E$ of every star.

\subsection{Input Catalog}
\label{sec:input-catalog}

\begin{figure*}[ht!]
\plotone{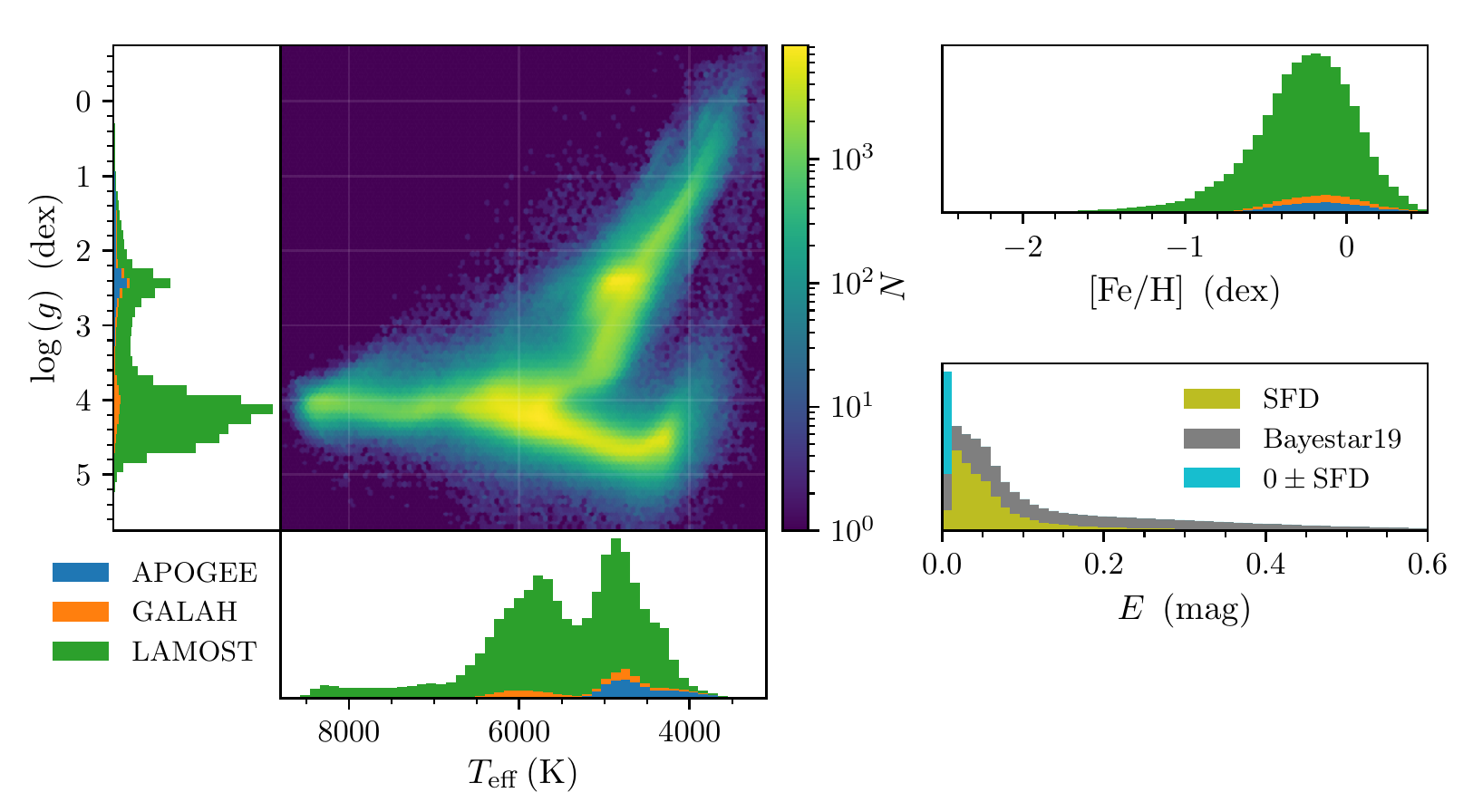}
  \caption{Distribution of the training data in $\teff$, $\logg$, $\feh$ and $E$. The spectroscopic parameter pipelines of LAMOST, APOGEE and GALAH are focused on analyzing relatively cool stars without strong molecular features, which restricts our coverage to $4000\,\mathrm{K} \lesssim \teff \lesssim 7500\,\mathrm{K}$. This range nevertheless encloses the majority of stars in magnitude-limited surveys. \label{fig:input-coverage}}
\end{figure*}

We construct our input catalog by matching each spectroscopic survey separately to the Gaia DR2 catalog, using a $0.2^{\prime\prime}$ matching radius. We require Gaia $G$ band photometry, and beyond that, opportunistically use whichever Gaia $BP$/$RP$, PS1, 2MASS and unWISE photometry meets our quality cuts for each star. We perform the catalog matching with the Large Survey Database \citep{Juric2012LSD}, using a radius of $0.2^{\prime\prime}$ to match PS1 to Gaia, and a radius of $0.4^{\prime\prime}$ to match 2MASS and unWISE to Gaia. When available, we include high-quality Gaia parallax measurements in the catalog (as described in \S\ref{sec:photometry-parallax}). For each source, we include the reddening estimates described in \S\ref{sec:reddenings}. We thus obtain one input catalog for each spectroscopic survey, which we stack to create one catalog. This results in a small number of repeated stars (39,580, or 1.4\% of the total), which we expect to have negligible impact on our results. Our resulting dataset has 2,888,361 stars. The number of stars with spectroscopy from each survey and photometry in each passband is given in Table \ref{tab:input-catalog}. Fig.~\ref{fig:input-coverage} shows the distribution of training data in stellar parameter and reddening space.

\begin{deluxetable*}{cccc}
    \tablecaption{The number of stars with spectroscopy from the given source or photometry in the given passband. \label{tab:input-catalog}}
    \tablehead{
    & \colhead{Source} & \colhead{N} & \colhead{\% of total}
    }
    \startdata
                 & LAMOST & 2,586,577 & 90\% \\
    Spectroscopy & GALAH & 122,411 & 4\% \\
                 & APOGEE & 179,373 & 6\% \\
    \hline
    Gaia parallaxes & $\varpi$ & 2,442,856 & 85\% \\
    \hline
                 & $G$ & 2,888,361 & 100\% \\
    Gaia         & $BP$ & 2,884,921 & 99.9\% \\
                 & $RP$ & 2,885,425 & 99.9\% \\
    \hline
                 & $g_{\mathrm{P1}}$ & 2,360,242 & 82\% \\
                 & $r_{\mathrm{P1}}$ & 2,032,495 & 70\% \\
    PS1          & $i_{\mathrm{P1}}$ & 1,853,962 & 64\% \\
                 & $z_{\mathrm{P1}}$ & 2,121,541 & 74\% \\
                 & $y_{\mathrm{P1}}$ & 2,428,790 & 84\% \\
    \hline
                 & $J$ & 2,742,702 & 95\% \\
    2MASS        & $H$ & 2,687,149 & 93\% \\
                 & $K_s$ & 2,576,599 & 89\% \\
    \hline
    unWISE       & $W_1$ & 839,532 & 29\% \\
                 & $W_2$ & 1,857,203 & 64\% \\
    \enddata
\end{deluxetable*}

\section{Training}
\label{sec:training}

We split our input catalog into three parts by randomly assigning sources to the training set (70\% of sources), validation set (20\% of sources) or test set (10\% of sources). We minimize the $\chi^2$ of our model on our training set, using the validation set to track progress of our training and to diagnose possible overfitting. All results, goodness-of-fit measures and validation plots in the following use the test set.

We implement our model in Tensorflow 2 \citep{tensorflow2015-whitepaper} with Keras \citep{chollet2015keras}, using the Adam optimizer \citep{Kingma2014}. We train the model in 20 iterations, with each iteration consisting of 25 training epochs. In the first iteration, we employ a learning rate of 0.001, and assume that $\nabla_{\vec{\theta}} \vec{M} = 0$ and $\vec{R} = 0$ when calculating the covariance matrix $C_m$ (see Eq.~\ref{eqn:covariance-matrix}). This means that in the initial iteration, uncertainties in the spectroscopic features and reddening do not propagate into the covariance matrix of the photometry. In order to minimize the impact of the missing reddening uncertainty term in the covariance, we exclude stars with $\sigma_E > 0.2$ from the initial training iteration. After each iteration, we multiply the learning rate by a factor of $e^{-\nicefrac{1}{5}}$. We use the trained network to update our estimate of the reddening of each source (see Eqs.~\ref{eqn:E-estimate}~--~\ref{eqn:sigma_E-estimate}) and to recalculate the terms in the covariance matrix that depend on $\vec{M}$, $\vec{R}$ and $\hat{E}$. Because $\vec{M} \args{\vec{\theta}\,}$ and $\vec{R} \args{\vec{\theta}\,}$ are represented by neural networks, obtaining the gradient terms in the covariance matrix, $\nabla_{\vec{\theta}} \vec{M}$ and $\nabla_{\vec{\theta}} \vec{R}$, is trivial. Finally, after each iteration, we calculate the $\chi^2 / \mathrm{d.o.f.}$ of each star. Because we re-estimate the reddening of each star, the number of degrees of freedom is one less than the number of observed passbands. We exclude stars that fail a cut from the next iteration. The threshold of the cut is reduced each iteration, beginning at $\chi^2 / \mathrm{d.o.f.} > 100$ after the first iteration, and decreasing logarithmically each iteration, such that it reaches $\chi^2 / \mathrm{d.o.f.} > 5$ by the $15^{\mathrm{th}}$ iteration, after which the threshold is held constant. In the final iteration, 2.7\% of stars are excluded based on this cut.

We check for over-fitting by comparing the loss on the training and validation datasets. A model is over-fit if it has learned peculiarities of the training dataset, rather than only the features that generalize to new data that the model was not trained on. Thus, over-fitting generally causes the loss obtained on the training dataset to be significantly lower than the loss obtained on the validation dataset. In our final iteration, the losses are 0.501 and 0.502 on the training and validation datasets, respectively, indicating that our model does not over-fit the training data. The contribution of the regularization terms (see Eq.~\ref{eqn:loss-with-regularization}) to this loss is small, so that the loss is approximately equal to $\langle \chi^2 / n_{\mathrm{bands}} \rangle$ (where the average is taken over stars). Because photometric bands that are not observed do not contribute to $\chi^2$, but are counted in $n_{\mathrm{bands}}$, the loss function is significantly less than unity. Taking into account only observed passbands, and disregarding stars with $\chi^2 / \mathrm{d.o.f.} > 5$, we find that $\langle \chi^2 / \mathrm{d.o.f.} \rangle = 0.87$ on the test dataset. This is slightly less than the expected value of unity, and likely indicates that some of the error sources (photometric, spectroscopic, parallax or reddening uncertainties) in our input catalog are systematically overestimated.

\section{Results}
\label{sec:results}

\begin{figure*}[ht!]
\plotone{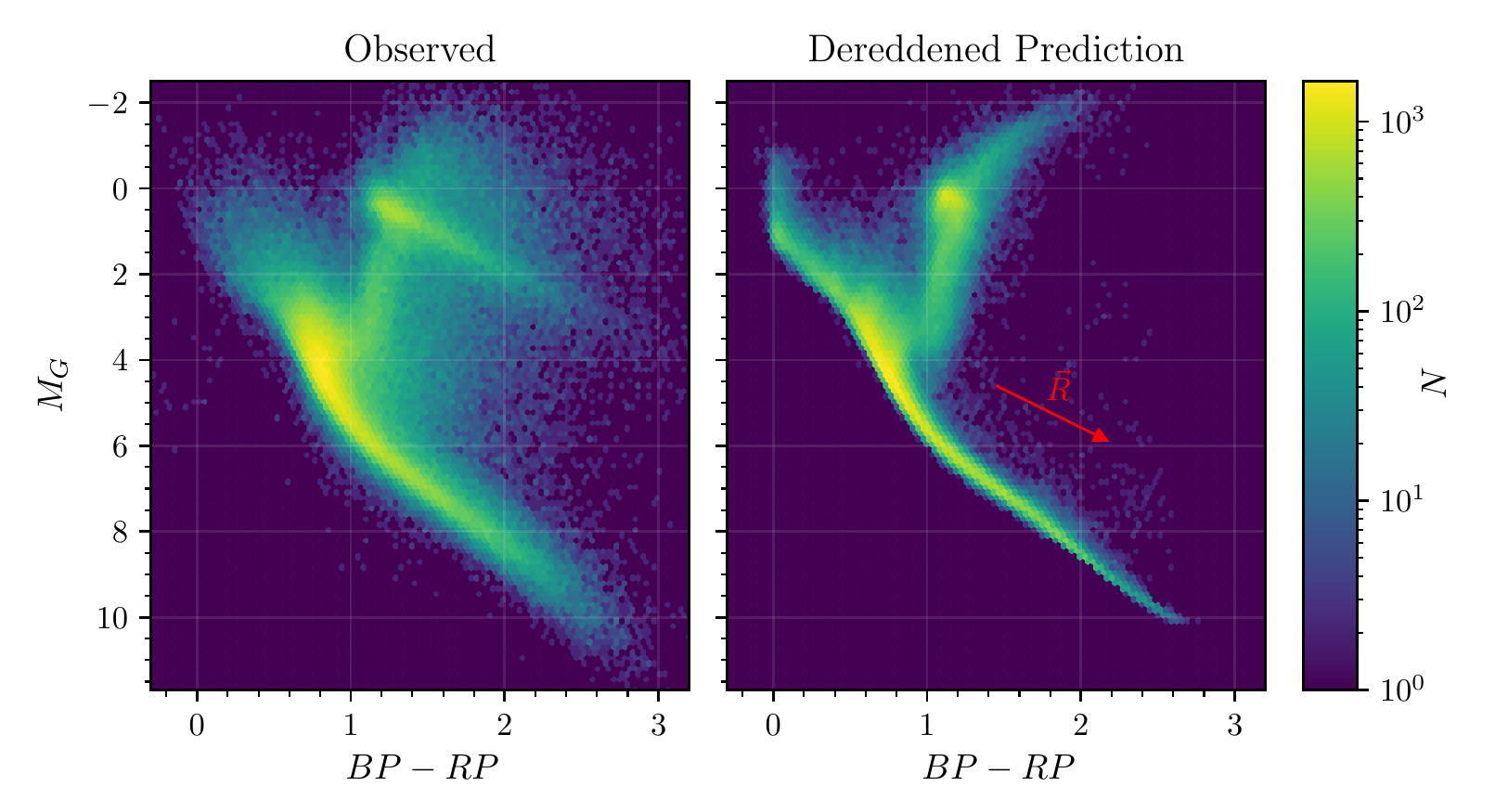}
  \caption{Gaia color-magnitude diagrams of observed (left) and predicted zero-reddening (right) photometry. The average learned reddening vector for the Gaia filters is shown in red. \label{fig:color-mag-density}}
\end{figure*}

Our trained model encodes both the mapping ${f \, : \, \vec{\theta} \, \mapsto \vec{M}}$, from the spectroscopic features ($\teff$, $\logg$ and $\feh$) to absolute magnitudes in 13 passbands, and the extinction vector $\vec{R}$ (including a weak dependence on the spectroscopic features). The left panel of Fig.~\ref{fig:color-mag-density} shows a Gaia color-magnitude diagram of the observed test-dataset photometry. We can map the measured spectral types of the stars in the test dataset, $\hat{\theta}$, to unextinguished absolute magnitudes, to show what the color-magnitude diagram would look like in the absence of dust and photometric errors. The right panel of Fig.~\ref{fig:color-mag-density} shows this ideal color-magnitude diagram, as well as the ``average'' extinction vector (obtained using the mean spectroscopic features in our dataset).

\begin{figure*}[ht!]
\plotone{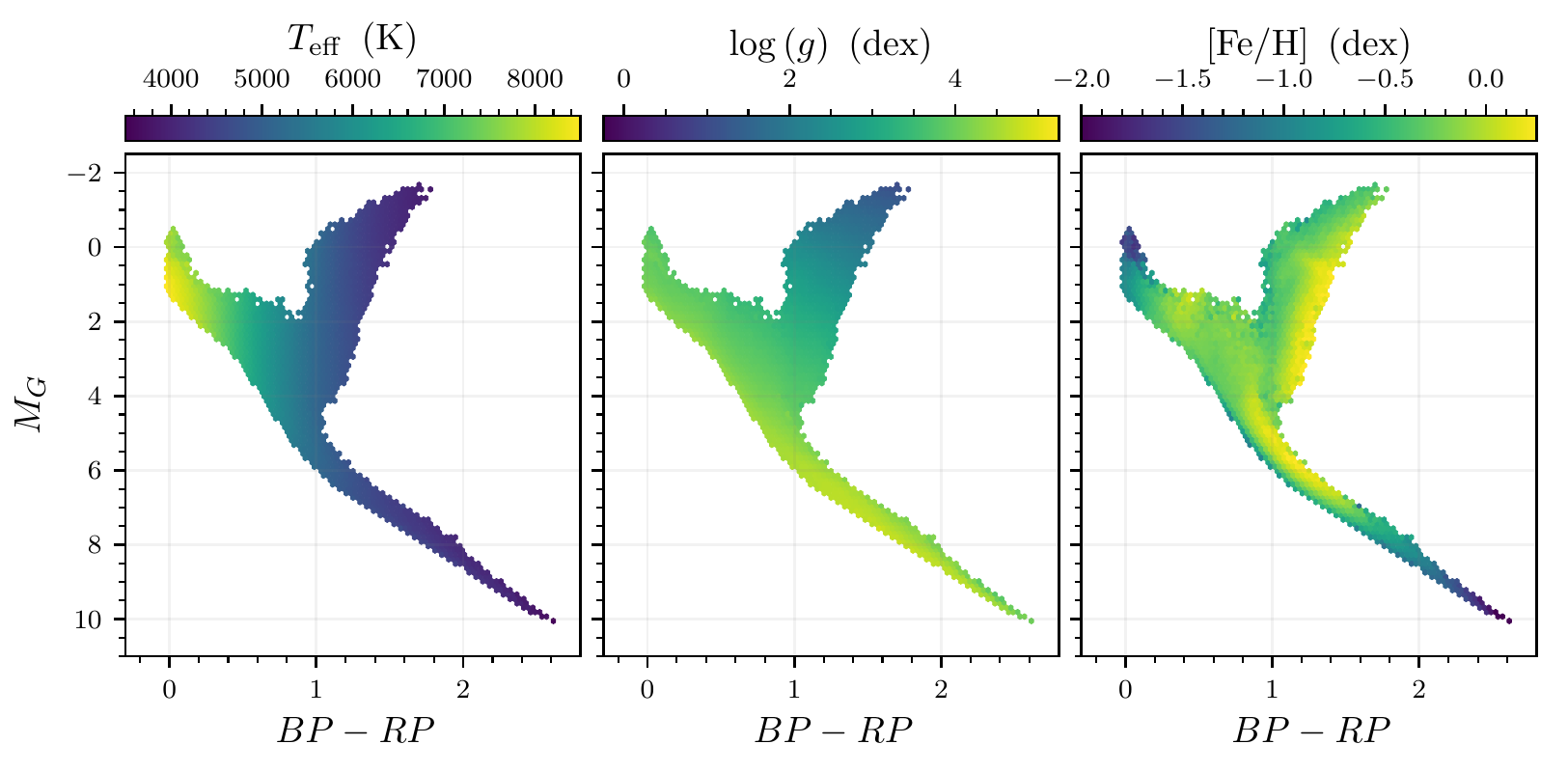}
  \caption{The mean spectroscopic features of stars in the test dataset, as a function of their predicted (rather than observed) dereddened photometry in color-magnitude space. This is one way to visualize the mapping from spectroscopic features to photometry. \label{fig:color-mag-theta}}
\end{figure*}

In Fig.~\ref{fig:color-mag-theta}, we visualize the mapping from spectroscopic features to color-magnitude space. We predict the zero-extinction absolute magnitudes of each star in the test dataset, and plot the mean spectroscopic features of stars falling in each region of color-magnitude space. These predictions have reduced aleatoric errors compared to the observations. In particular, they do not contain the final three error terms in Eq.~\eqref{eqn:model-with-errors}: photometric errors, error from the reddening estimate, and parallax errors. The only aleatoric error term which is present is the error propagated from the spectroscopic features to the absolute magnitudes:
\begin{align}
  \left(
    \delta \vec{\theta} \cdot
    \nabla_{\vec{\theta}}
  \right) \!\!
  \left.
    \vec{M}
  \right|_{\vec{\theta} = \hat{\theta}}
  \, .
\end{align}
By construction, however, this error term does not move a star's photometry off of the stellar locus. It moves each star in a random direction \textit{along} the stellar locus, but an ensemble of stars plotted in this manner will still fall along the modeled stellar locus. The epistemic errors deriving from any inaccuracies in our model itself remain in these reduced-error predictions.

\begin{figure*}[ht!]
  \plottwo{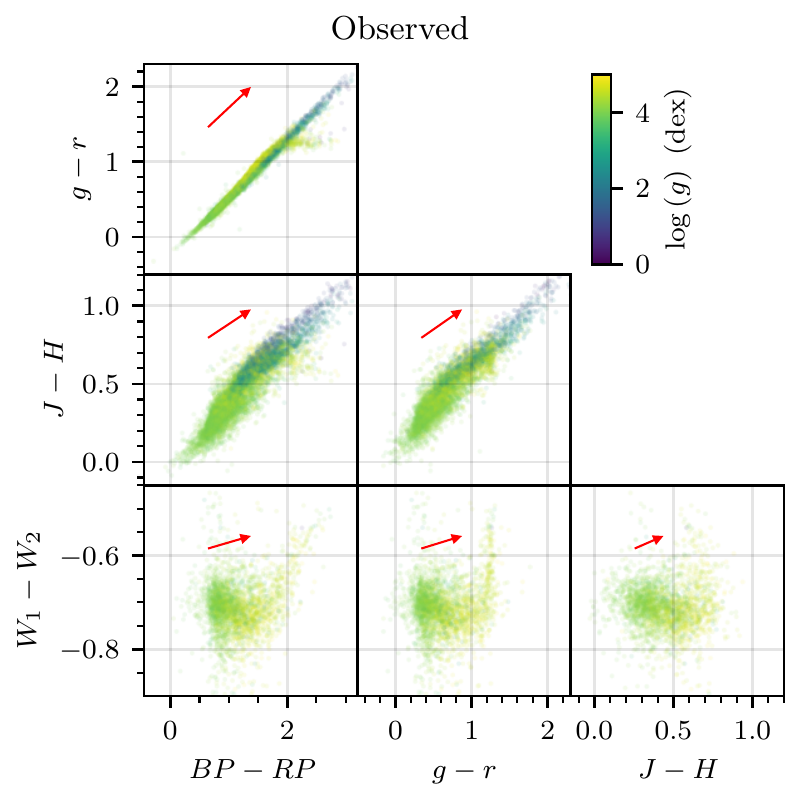}{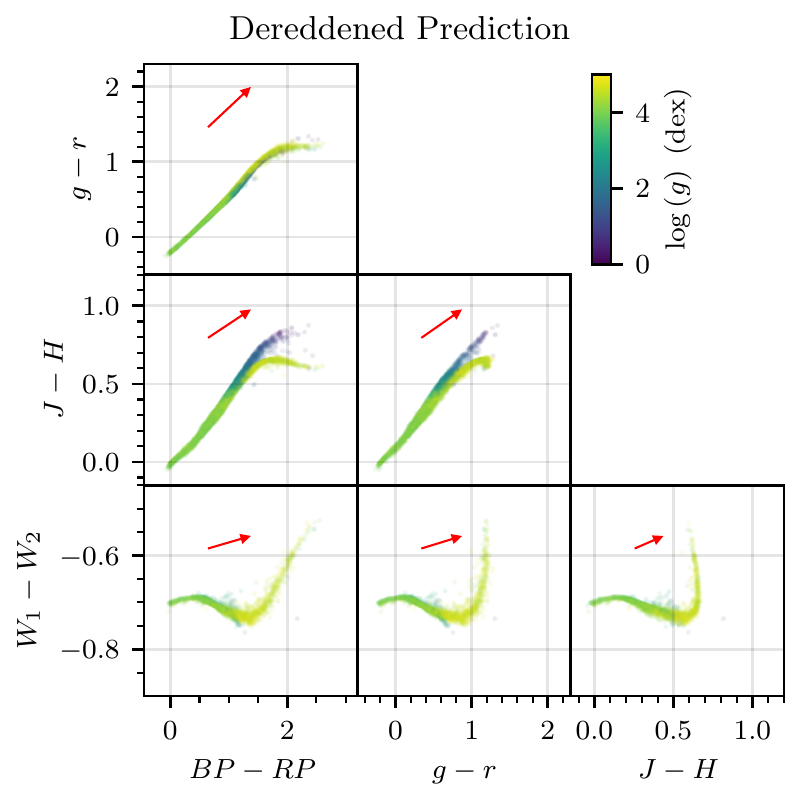}
  \caption{Color-color diagrams of observed (left) and predicted dereddened (right) photometry of a random subset of stars in the test dataset. The stars are colored by their measured $\log \left(g\right)$. The red arrows represent the learned reddening vector. The observations in the left panel include the effects of reddening and photometric noise, while the right panels show the predicted zero-reddening photometry without observational errors. \label{fig:color-color-logg}}
\end{figure*}

\begin{figure*}[ht!]
  \plottwo{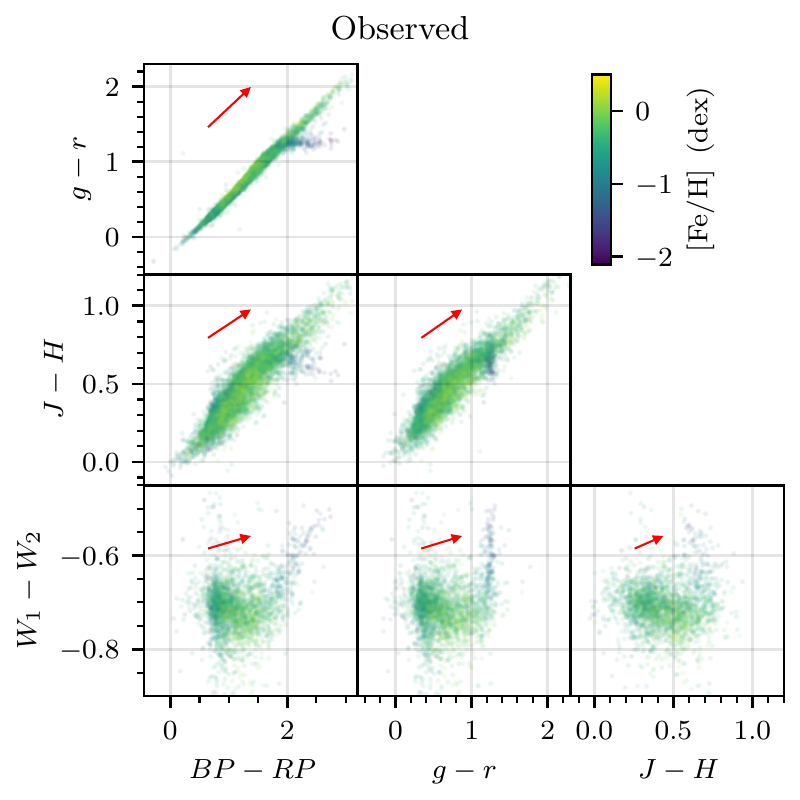}{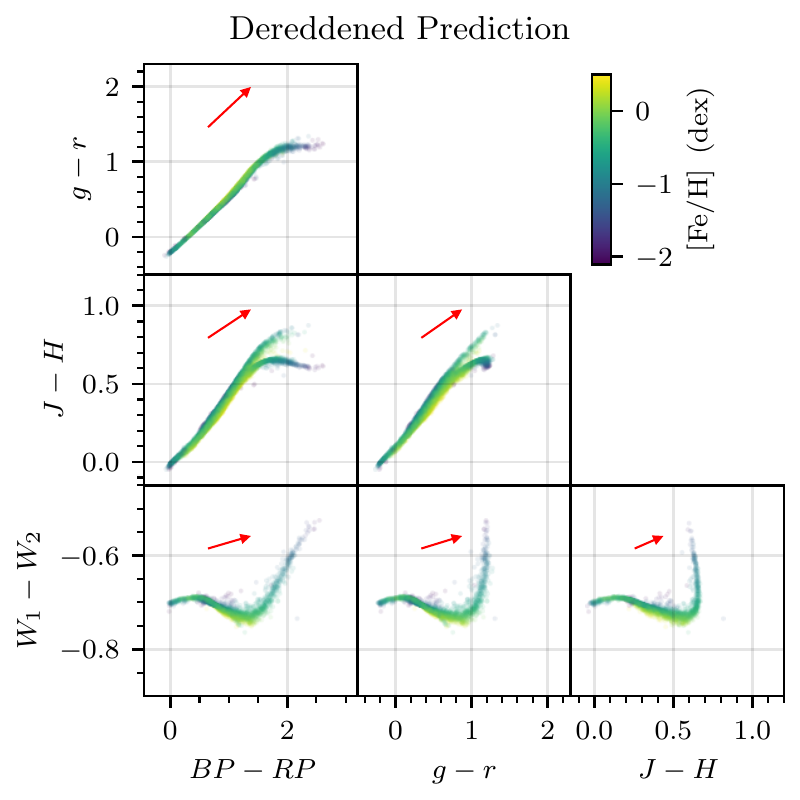}
  \caption{As Fig.~\ref{fig:color-color-logg}, but colored by measured $\feh$. \label{fig:color-color-feh}}
\end{figure*}

By leveraging observations of hundreds of thousands of stars, our model is able to learn the shape of the de-noised stellar locus with far greater precision than the typical photometric uncertainties of the input data. This can be illustrated by comparing observed, noisy photometry with predicted zero-extinction, de-noised photometry. The left panel of Fig.~\ref{fig:color-color-logg} shows observed test-set photometry, colored by $\logg$. In the right panel, we use the observed spectroscopic features $\hat{\theta}$ of each star to predict its photometry. The projected reddening vector is plotted for each combination of colors. Fig.~\ref{fig:color-color-feh} shows the same results, colored by $\feh$. We recover color trends with $\feh$ that are not immediately apparent in the noisy observational data, with more metal-rich stars being slightly redder in $g_{\mathrm{P1}} - r_{\mathrm{P1}}$ than metal-poor stars. This color trend with metallicity can also be seen in the right panel of Fig.~\ref{fig:color-mag-theta}.

\begin{figure}[ht!]
  \epsscale{1.1}
  \plotone{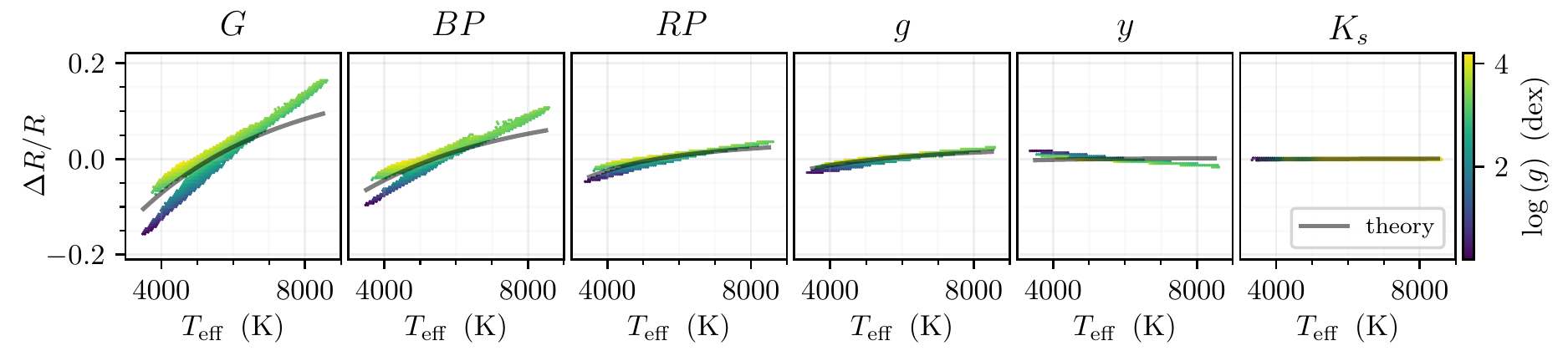}
  \caption{Dependence of our inferred reddening vector, $\vec{R}$, on the spectroscopic features in a subset of the passbands used in this study. Even for a universal extinction curve $\vec{A} \left(\lambda\right)$, the passband-integrated extinction, reflected in $\vec{R}$, will depend on the shape of the stellar spectrum, set primarily by $\teff$. Therefore, we show variations in $\vec{R}$ as a function of $\teff$, plotted on the $x$-axis. There is a weaker dependence on $\logg$. For each star in the test dataset, we calculate $\vec{R} (\hat{\theta})$. The $y$-axis shows the fractional deviation of each stellar $R$ from the value calculated for the mean spectroscopic features in our input catalog (see Eq.~\ref{eqn:dR-over-R}). Each pixel on the plot is colored by the mean $\logg$ of stars falling into the pixel. Pixels with fewer than 10 stars are colored white. The gray curve shows the theoretical result, using a blackbody source spectrum, measured transmission curves and the F99 extinction relation. \label{fig:dR-over-R}}
\end{figure}

In Fig.~\ref{fig:dR-over-R}, we show the dependence of our inferred extinction vector, $\vec{R}$ on the spectroscopic features for a subset of photometric passbands used in this work. On the $y$-axis, we plot the fractional variation of the components of the reddening vector,
\begin{align}
    \frac{\Delta R}{R}
    \equiv
    \frac{
        R (\vec{\theta}\,) - R_0
    }{
        R_0
    }
    \, ,
    \label{eqn:dR-over-R}
\end{align}
where $R_0$ is the value of $R$ calculated at the mean spectroscopic features in our input catalog. The $x$-axis shows $\teff$. Each pixel in the plot is colored by the average $\logg$ of stars falling in the pixel. We overplot the theoretical result obtained using a blackbody spectrum, measured transmission curves for each passband \citep{ApellanizWeiler2018,PS1PhotometricSystem,Cohen2003,Wright2010,SVO1,SVO2}, and the \citet[][``F99'']{Fitzpatrick1999} extinction relation. As expected, the variation in the reddening vector is primarily dependent on $\teff$, as temperature has the largest effect on the overall shape of the stellar spectrum. The reddening vector varies the most (by $\pm 15\%$) in Gaia $G$, due to the width of the passband (see Appendix~\ref{app:nonlinear-extinction}). The variation in the reddening vector is similar to theoretical expectations in the bluest passbands, where the variation is strongest: $G$, $BP$, $RP$ and $g_{\mathrm{P1}}$. However, our model learns a slight negative trend in $\Delta R / R$ with temperature in $i_{\mathrm{P1}}$, $z_{\mathrm{P1}}$ and $y_{\mathrm{P1}}$, which is unphysical, and likely due to unmodeled systematics in the training data. Our model also learns a slight difference in the reddening vector for dwarfs and giants, which manifests itself as a dependence of $\vec{R}$ on $\logg$, and which may be unphysical. This effect is also greatest in Gaia $G$, where $R$ differs by $\sim$7\% for 4000~K dwarfs and giants. The unphysical relations learned by the model can be suppressed with stronger L1 weight regularization in the extinction model. By contrast, the strong trends in $\Delta R / R$ with temperature that our model learns for $G$ and $BP$ -- which roughly match theoretical expectations -- are robust even under far stronger regularization.

\begin{figure}[ht!]
  \epsscale{0.7}
  \plotone{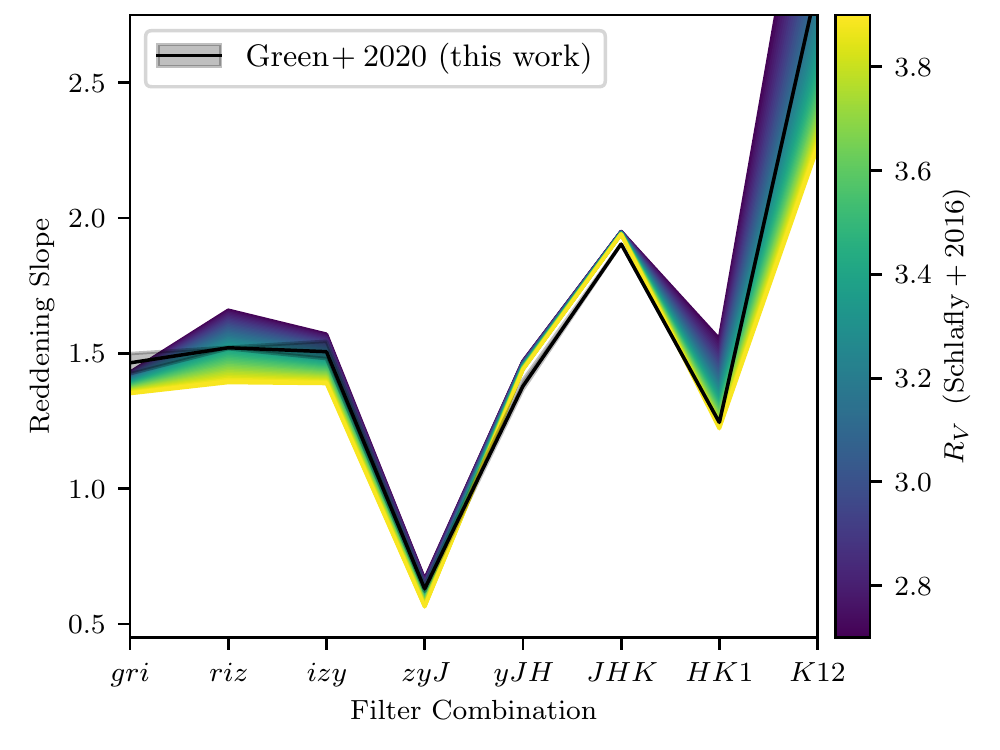}
  \caption{Our inferred reddening vector, compared to the family of reddening vectors found in \citet{Schlafly2016}. The $y$-axis, the ``reddening slope,'' represents the ratio of the reddening in one set of bands to the reddening in another set of bands, while the $x$-axis shows the set of bands chosen. For example, the $y$-value at $gri$ is equal to ${ E\left(g_{\mathrm{P1}}-r_{\mathrm{P1}}\right) / E\left(r_{\mathrm{P1}}-i_{\mathrm{P1}}\right) }$. For compactness, we abbreviate $K_s$, $W_1$ and $W_2$ as $K$, 1 and 2, respectively. We show the \citet{Schlafly2016} reddening vector for a range of $R_V$ values. Our inferred reddening vector depends weakly on the spectroscopic features, and the gray band shows the $1\sigma$ range of predicted reddening vectors in our test dataset. \label{fig:reddening-slopes}}
\end{figure}

We can also compare our inferred extinction vector to previous results. In Fig.~\ref{fig:reddening-slopes}, we compare our extinction vector to the family of vectors derived in \citet[][``S16'']{Schlafly2016}. S16 works in color-space, rather than with absolute magnitudes, and therefore derives the reddening -- as opposed to extinction -- vector. We therefore compare the slope of our inferred reddening curve, defined as the ratio of reddening in subsequent colors, such as ${ E\left(g_{\mathrm{P1}}-r_{\mathrm{P1}}\right) / E\left(r_{\mathrm{P1}}-i_{\mathrm{P1}}\right) }$. In most color combinations, the slope of our reddening vector falls close to that of the ``mean'' ($R_V \approx 3.3$) S16 reddening vector. However, we do see a somewhat different slope in two color combinations, ${ E\left(g_{\mathrm{P1}}-r_{\mathrm{P1}}\right) / E\left(r_{\mathrm{P1}}-i_{\mathrm{P1}}\right) }$ and ${ E\left(y_{\mathrm{P1}}-J\right) / E\left(J-K_{s}\right) }$.

Our trained neural network model and a tutorial on how to use it are available at \href{https://doi.org/10.5281/zenodo.3902382}{DOI:10.5281/zenodo.3902382}.

\section{Validation}
\label{sec:validation}

In order to validate the performance of our model in explaining the data, we investigate the behavior of residuals in predicted stellar colors in the test dataset, as a function of the spectroscopic features and reddening. For each color, $X-Y$, we consider the score of the prediction:
\begin{align}
    \frac{\Delta \left( X-Y \right)}{\sigma}
    &\equiv
    \frac{
        \left( X-Y \right)_{\mathrm{observed}} - \left( X-Y \right)_{\mathrm{predicted}}
    }{
        \sigma_{\left(X-Y\right)}
    }
    \,,
\end{align}
where $\sigma_{\left(X-Y\right)}$ is determined from the covariance matrix $C_m$ (see Eq.~\ref{eqn:covariance-matrix}), and the predicted colors are determined from the observed spectroscopic features and estimated reddening using the trained model and learned extinction vector: ${\vec{M} (\hat{\theta}) + \hat{E} \vec{R} (\hat{\theta})}$. We use the best-fit reddening estimates for this comparison, determined according to Eqs.~\ref{eqn:E-estimate}~--~\ref{eqn:sigma_E-estimate}.

\begin{figure*}[ht!]
  \plotone{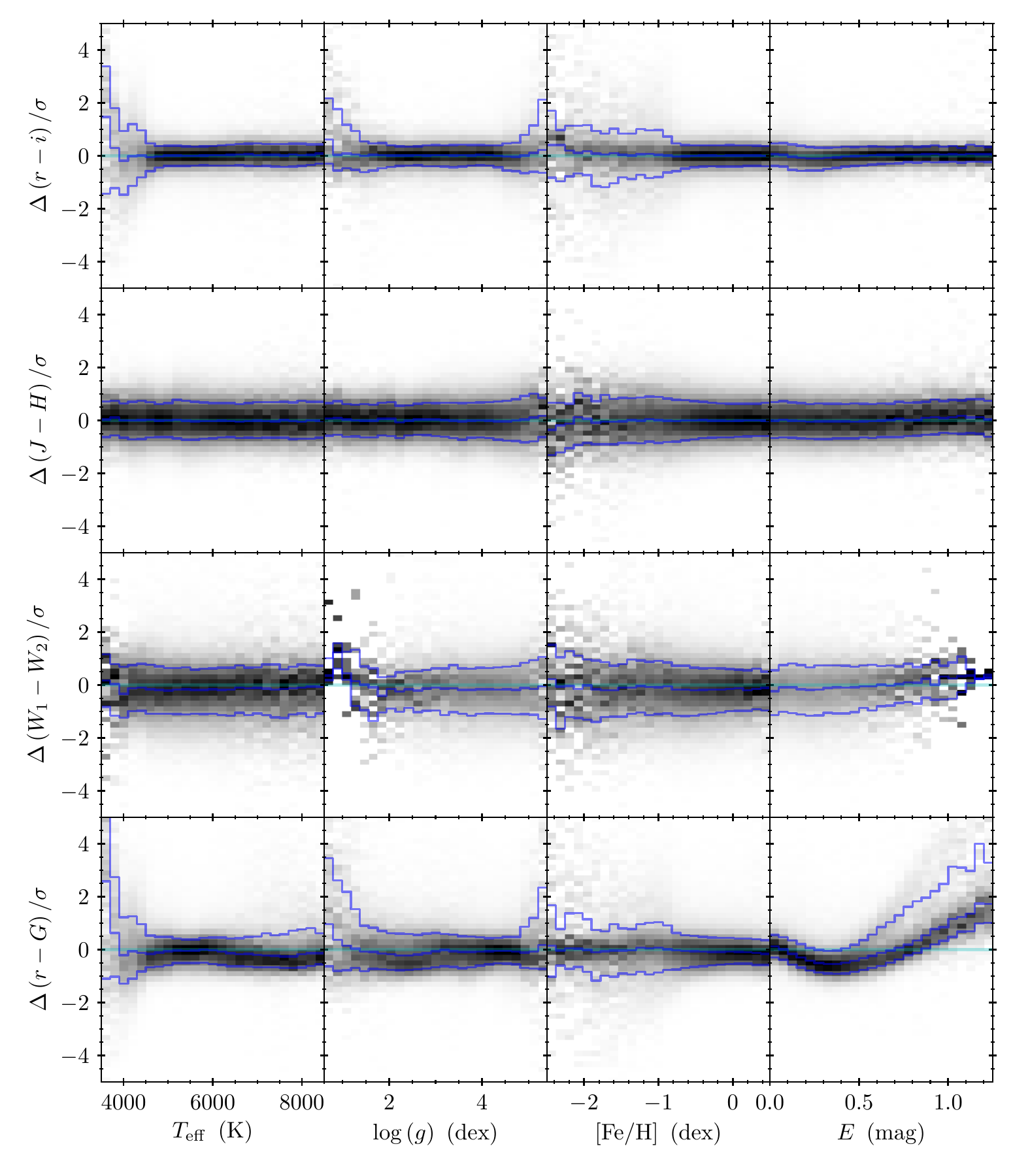}
  \caption{Distribution of color residuals (predicted minus observed, divided by the standard deviation) as a function of $\teff$, $\logg$, $\feh$ and $E$, for a subset of colors. The blue lines show the $16^{\mathrm{th}}$, $50^{\mathrm{th}}$ and $84^{\mathrm{th}}$ percentiles. Residuals for colors that do not involve Gaia $G$ show negligible trends, except at the extremes of the input parameter ranges. Colors involving Gaia $G$ exhibit significant residual trends, particularly as a function of reddening, possibly due to non-linear extinction effects caused by the exceptionally large spectral width of the $G$ passband. \label{fig:residual-trends}}
\end{figure*}

Fig.~\ref{fig:residual-trends} shows the distribution of these scores for four colors, as a function of the spectroscopic features and estimated reddening. For colors that do not involve the Gaia $G$ passband, the median residuals are close to zero, and there are no large trends as a function of the spectroscopic features or reddening. However, we observe trends in the median residuals of the ${r_{\mathrm{P1}}-G}$ color with reddening, $E$. Similar trends are observed in all colors involving Gaia $G$. These trends are presumably due to the width of the Gaia $G$ passband. In particular, our assumption that extinction is described by a vector $\vec{R} (\vec{\theta}\,)$ breaks down for extremely wide passbands, as extinction depends \textit{non-linearly} on the column density of dust. This effect is described in more detail in Appendix~\ref{app:nonlinear-extinction}.

For most colors, the $16^{\mathrm{th}}$ and $84^{\mathrm{th}}$ percentiles of the scores approximately span the range $\pm1$, as expected. In some colors, such as ${g_{\mathrm{P1}}-r_{\mathrm{P1}}}$, the scores are closer to zero than expected, indicating that our uncertainties are over-estimated. This could be due to the 0.02~mag floor we place on photometric uncertainties, to our floor on reddening uncertainty, or to the fact that we use diagonal covariance matrices for the spectroscopic features.

At low temperatures (${\teff \lesssim 4500\,\mathrm{K}}$) and surface gravities (${\logg \lesssim 2\,\mathrm{dex}}$), the ${g_{\mathrm{P1}}-r_{\mathrm{P1}}}$ residuals become larger, possibly due to decreased reliability in the spectroscopic features in this regime. In particular, line-blanketing by molecular species at low temperatures renders accurate determination of spectroscopic features more difficult.

\begin{figure*}[ht!]
  \plotone{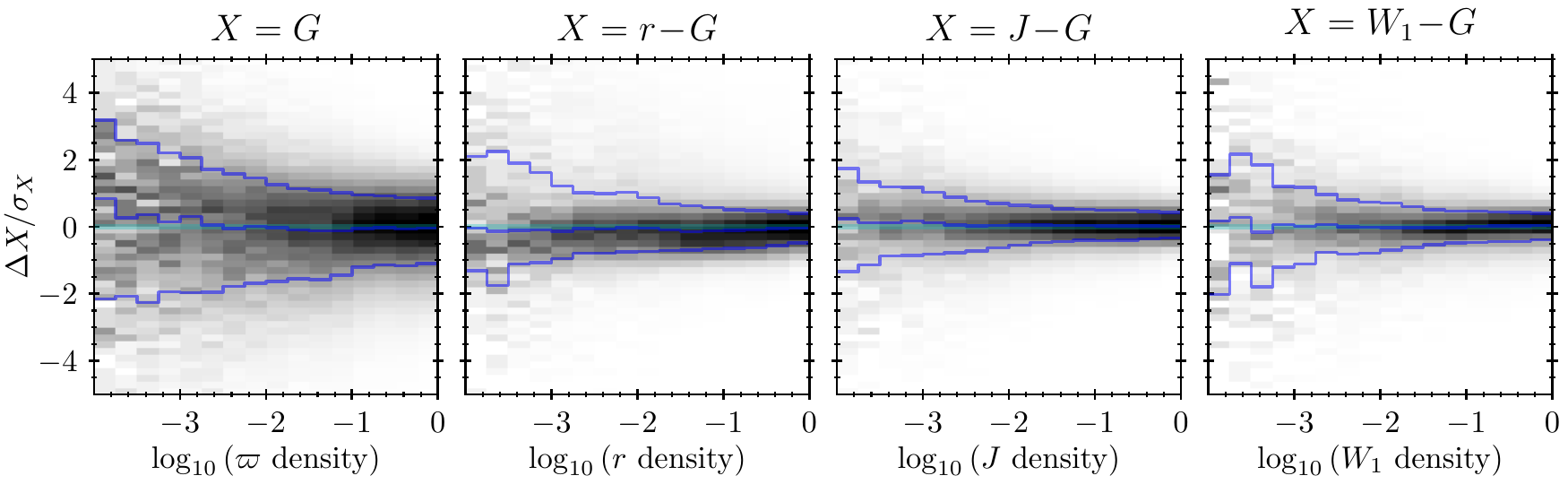}
  \caption{Distribution of photometric residuals (predicted minus observed, divided by the standard deviation), plotted on the $y$-axis, as a function of local density (in spectroscopic-feature-space) of the training data, plotted on the $x$-axis, for a subset of passbands. The blue lines show the $16^{\mathrm{th}}$, $50^{\mathrm{th}}$ and $84^{\mathrm{th}}$ percentiles. For $M_G$ (leftmost panel), we plot the density of parallax training data on the $x$-axis, as parallax observations are key to learning absolute magnitudes. In the right three panels, which show the residuals of $r-G$, $J-G$ and $W_1-G$ colors, we plot the density of $r$-, $J$- and $W_1$-band observations on the $x$-axis, respectively. All stars in our training catalog are observed in $G$-band. \label{fig:chi-vs-input-density}}
\end{figure*}

Our model is valid only in regions of $\vec{\theta}$-space in which we have training data. In particular, the spectroscopic pipelines from which we draw our features are tailored to stars with $4000\,\mathrm{K} \lesssim \teff \lesssim 7500\,\mathrm{K}$. The region covered by our training data also differs for each photometric passband, and depends on the overlap between photometric and spectroscopic surveys. We quantify the region of $\vec{\theta}$-space where our photometric predictions should be accurate, by measuring the density of stars in the input catalog. For each photometric passband, we construct a kernel density estimate of the input stars in $\vec{\theta}$-space, using a bandwidth of (50~K, 0.05~dex, 0.05~dex) in $\left( \teff, \logg, \feh \right)$. For each passband, we normalize the peak density to unity. Fig.~\ref{fig:chi-vs-input-density} shows the dependence of photometric residuals on this input density for a subset of passbands. As expected, photometric residuals are smallest in the regions of spectroscopic-feature-space with the highest density of training data. We find that our color predictions degrade more gracefully with increasing input density than our predictions of absolute $G$-band magnitude. One may set a density threshold below which one considers our model predictions unreliable for a given passband, based on the acceptable level of scatter in the photometric residuals. At an input density (for parallax observations) of $10^{-2}$, $\left< \left( \Delta M_G / \sigma_G \right)^2 \right> \sim 3$. Residuals in $X-G$ colors reach this level at input densities (in band $X$) of approximately $10^{-3}$. As a general rule of thumb, we therefore recommend to apply an input density threshold of $10^{-2}$ when absolute magnitudes are critical, and of $10^{-3}$ when only color predictions are required.

\section{Discussion}
\label{sec:discussion}

In this paper, we have learned a model that maps spectroscopic features ($\teff$, $\logg$ and $\feh$) to photometry, and at the same time describes the effect of dust on stellar photometry. Because this model is learned directly from the data, rather than using synthetic spectra and models of atmospheric, filter and instrument transmission, it sidesteps potential problems in these spectral and transmission models. With a large number of input stars, we are able to average down the statistical errors in the photometry, and learn stellar magnitudes to better than the typical photometric uncertainties of individual stars in our training data. Our method is not restricted to any specific set of photometric passbands, and can be applied to develop accurate mappings from spectroscopic features to absolute magnitudes for whichever set of passbands one wishes to use for a given study. The only requirement is that the new photometric survey overlap sufficiently with spectroscopic surveys, so that it is possible to construct a training dataset. Because our model has no concept of stellar spectra, filter transmission curves, atmospheric transmission and other physical effects that underly the mapping from spectroscopic features to photometry, it is unable to extrapolate to passbands that are not present in the training data. However, our model is not computationally expensive to train. Provided that training data is available, training a new model with additional passbands is a quick process.

Our trained model can be used to infer the spectroscopic features, distances and extinctions of large numbers of stars with observed photometry -- in the absence of spectral data. Using Bayes' theorem, the posterior density of $\vec{\theta}$, $\mu$ and $E$, given observed photometry $\hat{m}$, is given by
\begin{align}
    \pconds{\vec{\theta} ,\, \mu ,\, E}{\hat{m}}
    &=
    \frac{
        \pconds{\hat{m}}{\vec{\theta} ,\, \mu ,\, E} \,
        p \args{\vec{\theta} ,\, \mu ,\, E}
    }{
        p \arg{\hat{m}}
    }
    \, ,
\end{align}
where $p \arg{\hat{m}}$ is a normalizing constant. Because our model maps from spectroscopic features to absolute magnitudes and the reddening vector, $\vec{R}$, it can be used to construct an accurate likelihood function, $\pconds{\hat{m}}{\vec{\theta} ,\, \mu ,\, E}$. The above inference additionally requires a prior on the distribution of stars and dust throughout the Milky Way, ${p \args{\vec{\theta} ,\, \mu ,\, E}}$. There are many Galactic models that one could choose to construct such a prior, such as the Besan\c{c}on Model \citep{Robin2012} or the SDSS Tomography models \citep{Juric2008TomographyI,Ivezic2008TomographyII}. More ambitiously, the Galactic model itself could be inferred hierarchically, using the distribution of a large number of observed stars and detailed knowledge of the selection function. However, the problem of inferring stellar parameters from photometry exceeds the scope of this paper, which is instead focused on developing a highly accurate mapping from spectroscopic features to stellar photometry. One promising area for the application of our data-driven stellar model is three-dimensional dust mapping, which often relies on hundreds of millions of stellar distances and reddenings inferred from photometric surveys \citep{Green2019,Lallement2019,Chen2019,LeikeEnsslin2019,Andrae2018}.

In order to infer stellar parameters, we would like to construct models that cover as wide as possible a range of stellar parameters. With the increasing availability of spectroscopic data from ongoing surveys, such as LAMOST, APOGEE and GALAH, and from upcoming surveys such as SDSS-V \citep{Kollmeier2019}, it will be possible to train models that span a wider range of spectroscopic features.

Our model also provides a clean and accurate way to convert stellar evolutionary models to photometric predictions. Stellar evolutionary models begin with intrinsic stellar properties, such as initial mass and chemical abundances, and predict the evolution of stellar interior structure, abundances and atmospheric properties with age. In order to predict broad-band photometry, synthetic atmospheric models are typically used, often with empirical corrections to better match observed photometry. Our models provide a simple alternative -- and by construction accurate -- method of mapping from the spectroscopic features $\vec{\theta}$ output by stellar evolution models to stellar photometry.

Accurate stellar loci are also critical to developing color transformations between different photometric surveys. These color transformations are necessary for cross-calibration of photometric surveys \citep{Finkbeiner2016Hypercal,Scolnic2015Supercal}. For example, \citet[][``Hypercalibration'']{Finkbeiner2016Hypercal} recalibrates SDSS photometry using PS1, and develops an empirical color transformation between the two surveys that is applicable to main-sequence stars in low-reddening regions. This procedure encounters difficulties in SDSS $u$-band, however, due to the stronger metallicity dependence of stellar photometry in the near-ultraviolet. Accurate, metallicity-dependent color transformations derived using the methods presented in this work could therefore be of use in improving such cross-survey calibrations.

In this work, we have considered only the simplest possible extinction model, in which a single vector -- with weak dependence on the spectroscopic features -- describes the effect of dust on photometric magnitudes in the chosen passbands. However, the direction of this vector should depend on variation in the dust extinction spectrum itself, often parameterized by $R_V$ \citep{Fitzpatrick1999}. Here, we propose two possible avenues for future exploration of variation in the dust extinction spectrum. The first method is to model the residuals between the observed and predicted photometry, similarly to the analysis in \citet{Schlafly2016}. This is an after-the-fact analysis which can be applied to the residuals generated by the machinery developed in this paper. The second method is to build a model of the variation in the extinction vector directly into our neural network. This could be achieved by modeling the extinction of star $i$ as a sum of two or more components:
\begin{align}
    \vec{A}_i = \sum_{k=1}^{K} E_{ki} \vec{R}_k \, .
\end{align}
A separate dense neural network layer would represent each vector $\vec{R}_k$, while the extinction of each star $i$ would be described by $K$ coefficients, $E_{ki}$, $k = 1,\ldots K$. The vector $\vec{R}_1$ would represent a ``mean'' extinction relation, while $\vec{R}_2$ and higher-order vectors would describe variation in dust extinction properties. With appropriate regularizing conditions (e.g., orthogonality between the vectors $\big\{ \vec{R}_k \big\}$), such a model could be trained to obtain a more flexible and accurate description of dust extinction. Such a model could also be used as a basis to explore the spatial variation of dust properties throughout the Milky Way. We leave this extension of our model to future work.

\section{Conclusions}
\label{sec:conclusions}

In this work, we have presented a method for learning a mapping from spectroscopic features to absolute magnitudes. Our model additionally learns an extinction vector representing the effect of dust on stellar magnitudes in the chosen passbands. Our method can be easily adapted to an arbitrary set of photometric passbands or spectroscopic datasets, allowing inference of stellar parameters from a range of different photometric surveys.

We have demonstrated the effectiveness of our model on a dataset consisting of spectroscopic features from LAMOST, APOGEE and GALAH, parallaxes from Gaia, photometry from Gaia, PS1, 2MASS and WISE, and reddening estimates from SFD and Bayestar19. We obtain precision de-noised, zero-extinction stellar loci in the resulting 13-dimensional magnitude space, along with a model of the reddening vector $\vec{R}$ and its (weak) dependence on spectroscopic features.

We provide our trained neural network model, our estimate of its region of validity, and a tutorial on how to use the model at \href{https://doi.org/10.5281/zenodo.3902382}{DOI:10.5281/zenodo.3902382}.

This method is perhaps the simplest and most accurate way to convert stellar evolutionary models into photometric predictions. It also provides a simple method for constructing a highly accurate likelihood function of stellar photometry, given spectroscopic features, distance and extinction. This is a key component of methods which infer stellar parameters, distances and extinctions from photometric data. One immediate application of our method will be to three-dimensional dust mapping in the Milky Way, though we expect it to be of use in other fields, such as Galactic archaeology and stellar population synthesis.

\acknowledgments

The computations in this paper were run on the FASRC Cannon cluster supported by the FAS Division of Science Research Computing Group at Harvard University. Jan Rybizki, Morgan Fouesneau and Ren\'{e} Andrae were funded by the DLR (German space agency) via grant 50\,QG\,1403. This work is partly based on data from the European Space Agency (ESA) mission Gaia (\url{https://www.cosmos.esa.int/gaia}), processed by the Gaia Data Processing and Analysis Consortium (DPAC, \url{https://www.cosmos.esa.int/web/gaia/dpac/consortium}). Funding for the DPAC has been provided by national institutions, in particular the institutions participating in the Gaia Multilateral Agreement. 

\vspace{5mm}

\software{
  Astropy \citep{astropy2013,astropy2018},
  dustmaps \citep{Green2018dustmaps},
  Keras \citep{chollet2015keras},
  Large Survey Database \citep{Juric2012LSD},
  Matplotlib \citep{Hunter2007Matplotlib},
  NumPy \citep{NumPy},
  SciPy \citep{SciPy},
  Tensorflow 2 \citep{Abadi2016Tensorflow}
}

\appendix

\section{Converting Noisy Parallax to Distance Modulus}
\label{app:noisy-parallax}

In specifying our model, we calculate a mean distance modulus from a noisy measured parallax: $\mu \arg{\hat{\varpi}}$. As noted in the paper, we only attempt such a conversion when the parallax uncertainty is small. Here, we derive an estimate of the mean distance modulus in this regime.
\begin{align}
  \left< \mu \right>
    &= 10 - 5 \left<
        \log_{10} \left( \frac{\varpi}{1\,\mathrm{mas}} \right)
      \right>
      \\
    &= 10 - 5 \left[
        \log_{10} \left( \frac{\hat{\varpi}}{1\,\mathrm{mas}} \right)
        + \left< \log_{10} \left(
          1 + \frac{\varpi - \hat{\varpi}}{\hat{\varpi}}
        \right) \right>
      \right] \, .
    \label{eqn:mu-avg-expanded}
\end{align}
Defining
\begin{align}
  x \equiv \frac{\varpi - \hat{\varpi}}{\hat{\varpi}}
\end{align}
and Taylor-expanding the second logarithm in Eq.~\eqref{eqn:mu-avg-expanded} in terms of $x$, we obtain
\begin{align}
  \left< \mu \right>
    &= 10
      - 5 \log_{10} \left( \frac{\hat{\varpi}}{1\,\mathrm{mas}} \right)
      + \frac{5}{\ln 10} \sum_{n=1}^{\infty}
        \frac{\left( -1 \right)^n \left< x^n \right>}{n}
    \, .
    \label{eqn:mu-of-x}
\end{align}
The $n^{\mathrm{th}}$ moment of $x$ is given by
\begin{align}
    \left< x^n \right>
    &=
    \int_0^{\infty}
        \left( \frac{\varpi - \hat{\varpi}}{\hat{\varpi}} \right)^{\!\! n}
        \pcond{\varpi}{\hat{\varpi},\,\sigma_{\varpi}}
        \mathrm{d}\varpi
    \\
    &=
    \int_0^{\infty}
        \left( \frac{\varpi - \hat{\varpi}}{\hat{\varpi}} \right)^{\!\! n}
        \frac{
            \pcond{\hat{\varpi}}{\varpi,\,\sigma_{\varpi}}
            p\arg{\varpi}
        }{
            \pcond{\hat{\varpi}}{\sigma_{\varpi}}
        }
        \, \mathrm{d}\varpi
    \\
    &= \frac{K_n}{K_0}
    \, ,
\end{align}
where
\begin{align}
    K_n
    &\equiv
    \int_0^{\infty}
        \left( \frac{\varpi - \hat{\varpi}}{\hat{\varpi}} \right)^{\!\! n}
        \pcond{\hat{\varpi}}{\varpi,\,\sigma_{\varpi}}
        p\arg{\varpi}
        \, \mathrm{d}\varpi
    \, .
\end{align}
In order to evaluate $K_n$, we assume that the likelihood of $\hat{\varpi}$ is Gaussian, and we Taylor expand the prior, $p \arg{\varpi}$, around $\hat{\varpi}$. We additionally assume that $\hat{\varpi} \gg \sigma_{\varpi}$, so that we can extend the lower limit of the integrand to $-\infty$ with little change to the value of the integral. This turns $K_n$ into a sum over Gaussian integrals, which can be exactly evaluated:
\begin{align}
    K_n
    &\simeq
    \int_{-\infty}^{\infty}
        \left( \frac{\varpi - \hat{\varpi}}{\hat{\varpi}} \right)^{\!\! n}
        \mathcal{N} \arg{\hat{\varpi} \mid \varpi,\,\sigma_{\varpi}}
        \sum_{m=0}^{\infty}
            \frac{
                p^{\left(m\right)} \arg{\varpi=\hat{\varpi}}
            }{
                m!
            }
            \left( \varpi - \hat{\varpi} \right)^m
        \, \mathrm{d}\varpi
    \\
    &=
    \sum_{m=0}^{\infty}
        \frac{
            \hat{\varpi}^m
            p^{\left(m\right)} \arg{\varpi=\hat{\varpi}}
        }{
            m!
        }
    \int_{-\infty}^{\infty}
        \left( \frac{\varpi - \hat{\varpi}}{\hat{\varpi}} \right)^{\!\! n+m}
        \!\!\!\!\!\!
        \mathcal{N} \arg{\hat{\varpi} \mid \varpi,\,\sigma_{\varpi}}
        \, \mathrm{d}\varpi
    \\
    &=
    \sum_{\substack{{m=0}\\{n+m \, \mathrm{even}}}}^{\infty}
        \!\!\!
        \frac{
            \left(n\!+\!m\!-\!1\right)!!
        }{
            m!
        }
        \,
        \hat{\varpi}^m
        p^{\left(m\right)} \arg{\varpi=\hat{\varpi}}
        \left( \frac{\sigma_{\varpi}}{\hat{\varpi}} \right)^{\! n+m} \, .
\end{align}
To $2^{\mathrm{nd}}$ order in $\frac{\sigma_{\varpi}}{\hat{\varpi}}$,
\begin{align}
    K_0 &=
        p \arg{\varpi = \hat{\varpi}}
        + \frac{1}{2}
            \hat{\varpi}^2 p^{\prime\prime} \arg{\varpi = \hat{\varpi}}
            \left( \frac{\sigma_{\varpi}}{\hat{\varpi}} \right)^{\! 2}
    \, , \\
    K_1 &=
        \hat{\varpi} \, p^{\prime} \arg{\varpi = \hat{\varpi}}
        \left( \frac{\sigma_{\varpi}}{\hat{\varpi}} \right)^{\! 2}
    \, , \\
    K_2 &=
        p \arg{\varpi = \hat{\varpi}}
        \left( \frac{\sigma_{\varpi}}{\hat{\varpi}} \right)^{\! 2}
    \, .
\end{align}
Thus, the first two moments of $x$ are given by
\begin{align}
    \left< x \right>
        &= \frac{K_1}{K_0}
        \simeq
            \frac{
                \hat{\varpi} \, p^{\prime} \arg{\varpi = \hat{\varpi}}
            }{
                p \arg{\varpi = \hat{\varpi}}
            }
            \left( \frac{\sigma_{\varpi}}{\hat{\varpi}} \right)^{\! 2}
    \, , \\
    \left< x^2 \right>
        &= \frac{K_2}{K_0}
        \simeq
            \left( \frac{\sigma_{\varpi}}{\hat{\varpi}} \right)^{\! 2}
    \, .
\end{align}
All higher-order moments of $x$ depend on $\left( \frac{\sigma_{\varpi}}{\hat{\varpi}} \right)^{\! 4}$ or higher-order terms. Plugging the moments of $x$ into Eq.~\eqref{eqn:mu-of-x}, we obtain
\begin{align}
    \left< \mu \right>
    &\simeq
    10 - 5 \log_{10} \left( \frac{\hat{\varpi}}{1\,\mathrm{mas}} \right)
    +
    \frac{5}{\ln 10} \left( \frac{\sigma_{\varpi}}{\hat{\varpi}} \right)^{\! 2}
    \left[
        \frac{1}{2}
        -
        \frac{
            \hat{\varpi} \, p^{\prime} \arg{\varpi = \hat{\varpi}}
        }{
            p \arg{\varpi = \hat{\varpi}}
        }
    \right]
    +
    \mathcal{O} \! \left[
        \left( \frac{\sigma_{\varpi}}{\hat{\varpi}} \right)^{\! 4}
    \right]
    \, .
\end{align}
The lowest-order correction to the naive distance modulus thus depends on the slope of the prior on parallax at the location of the measured parallax. The sign of the correction depends on the choice of the parallax (or equivalently, of the distance) prior, and scales with $\left( \frac{\sigma_{\varpi}}{\hat{\varpi}} \right)^{\! 2}$. For the cutoff we make when calculating the distance modulus, $\hat{\varpi} > 5 \sigma_{\varpi}$, we can therefore expect the maximum correction on the order of several hundredths of a magnitude. As the sign of the correction depends on the exact choice of the distance prior, we do not attempt to correct for the effect in this work, and instead use the naive conversion from measured parallax to distance modulus. The errors incurred in this manner only affect absolute magnitudes, but not colors, which are independent of distance.

\section{Non-linearity of extinction vs. optical depth in Gaia \textit{G} passband}
\label{app:nonlinear-extinction}

In this work, we assume that the effect of interstellar extinction can be described by a nearly universal vector, $\vec{R}$, with only weak dependence on the spectroscopic features. As in Eq.~\eqref{eqn:model-no-errors}, the extinction of each star is proportional to this vector. This is an idealization that holds only when the wavelength-vs.-dust-opacity relation does not change, and in the limit that the passbands we are working with are extremely narrow. However, the Gaia $G$ passband covers a wide range of wavelengths, and therefore exhibits measurable non-linear extinction (vs. dust column density) and stellar-spectrum-dependent extinction. In the following, we show how these effects arise, and estimate the non-linear extinction term.

The AB magnitude measured by a charge-coupled device (CCD) is given by
\begin{align}
    m_{AB} = -2.5 \log_{10} \left[
        \frac{
            \int \left( h \nu \right)^{-1} f_{\nu} \arg{\nu} T \arg{\nu} \mathrm{d}\nu
        }{
            \int \left( h \nu \right)^{-1} \left( 3631 \, \mathrm{Jy} \right) T \arg{\nu} \mathrm{d}\nu
        }
    \right] \, ,
\end{align}
where $T \arg{\nu}$ is the transmission of the instrument (including the optics, filter and CCD, and depending on convention, possibly the atmosphere) as a function of frequency, $f_{\nu} \arg{\nu}$ is the spectral flux density of the source, and the factors of $\left( h \nu \right)^{-1}$ correspond to the fact that CCDs count photons, rather than measuring incident energy. In the presence of interstellar dust with optical depth $\tau \arg{\nu}$, the extinction is given by
\begin{align}
    A = -2.5 \log_{10} \left[
        \frac{
            \int_0^{\infty} \left( h \nu \right)^{-1} f_{\nu} \arg{\nu} e^{-\tau \, \arg{\nu}} T \arg{\nu} \mathrm{d}\nu
        }{
            \int_0^{\infty} \left( h \nu \right)^{-1} f_{\nu} \arg{\nu} T \arg{\nu} \mathrm{d}\nu
        }
    \right] \, ,
    \label{eqn:extinctinon-def}
\end{align}
where we are now using $f_{\nu} \arg{\nu}$ to denote the source spectrum \textit{in the absence of dust}. For an infinitely narrow passband, $T \arg{\nu} = \delta \arg{\nu - \nu_0}$, extinction is given by
\begin{align}
    A = \frac{2.5}{\ln 10} \, \tau \arg{\nu_0} \, .
\end{align}
In this limit, the vector $\vec{R}$, describing the relative amount of extinction in each passband, can be calculated directly from the opacity of the dust at the central frequencies of the passbands, and does not depend on the spectroscopic features.

In order to explore the effect of passband width on extinction, consider a Gaussian passband with width $\Delta \nu$:
\begin{align}
    T \arg{\nu} = e^{-\frac{1}{2} \left( \nu - \nu_0 \right)^2 / \Delta \nu^2} \, .
\end{align}
The number of photons detected by the CCD is then given by
\begin{align}
    N \arg{\sigma}
    &= \int_0^{\infty} \underbrace{
            \left( h \nu \right)^{-1} f_{\nu} \arg{\nu} e^{-\kappa \, \arg{\nu} \sigma}
        }_{
            \equiv n \, \arg{\nu}
        }
        e^{-\frac{1}{2} \left( \nu - \nu_0 \right)^2 / \Delta \nu^2} \mathrm{d}\nu
    \, ,
\end{align}
where $\kappa \arg{\nu}$ is the opacity of the dust, and $\sigma$ is the dust column density, such that optical depth is given by $\tau \arg{\nu} = \kappa \arg{\nu} \sigma$. Taylor-expanding $n \arg{\nu}$ around $\nu = \nu_0$, we obtain
\begin{align}
    N \arg{\sigma}
    &= \sum_{k=0}
        \frac{1}{k!}
        \frac{
            \mathrm{d}^k n \arg{\nu_0}
        }{
            \mathrm{d} \nu^k
        }
        \int_0^{\infty} \!\!\!\!\!
        \left( \nu - \nu_0 \right)^k
        e^{-\frac{1}{2} \left( \nu - \nu_0 \right)^2 / \Delta \nu^2}
        \mathrm{d}\nu
    \, .
\end{align}
Extending the lower limit of the integral to $-\infty$ allows us to evaluate it exactly, yielding
\begin{align}
    N \arg{\sigma}
    &\simeq \sqrt{2 \pi} \, \Delta \nu \,
        \sum_{k \ \mathrm{even}}
            \frac{1}{k!!}
            \left( \Delta \nu \right)^k
            \frac{
                \mathrm{d}^k n \arg{\nu_0}
            }{
                \mathrm{d} \nu^k
            }
    \, ,
\end{align}
where $k!!$ is the double factorial of $k$. As long as $\Delta \nu \ll \nu_0$, changing the lower limit of the integral impacts the result minimally. The extinction is given by
\begin{align}
    A \arg{\sigma}
    &= -2.5 \log_{10} \left[ \frac{N \arg{\sigma}}{N \arg{\sigma = 0}} \right] \\
    &= -2.5 \log_{10} \left[
        \frac{
            n \arg{\nu_0, \sigma} + \frac{1}{2} \! \left(\Delta \nu\right)^2 \frac{\mathrm{d}^2 n \,\arg{\nu_0, \sigma}}{\mathrm{d}\nu^2} + \ldots
        }{
            n \arg{\nu_0, 0} + \frac{1}{2} \! \left(\Delta \nu\right)^2 \frac{\mathrm{d}^2 n \,\arg{\nu_0, 0}}{\mathrm{d}\nu^2} + \ldots
        }
    \right]
    \\
    &= \frac{2.5}{\ln 10} \, \kappa \arg{\nu_0} \sigma
        -
        2.5 \log_{10} \left\{
            1 + \frac{1}{2} \! \left(\Delta \nu\right)^2
            \left[
                \frac{\mathrm{d}^2 \ln n \,\arg{\nu_0, \sigma}}{\mathrm{d}\nu^2}
                -
                \frac{\mathrm{d}^2 \ln n \,\arg{\nu_0, 0}}{\mathrm{d}\nu^2}
            \right]
            + \mathcal{O} \arg{\Delta \nu^4}
    \right\}
    \\
    &\simeq
        \frac{2.5}{\ln 10} \kappa \arg{\nu_0} \sigma
        -
        \frac{1.25}{\ln 10}
        \left(\Delta \nu\right)^2
        \left[
            \frac{\mathrm{d}^2 \ln n \,\arg{\nu_0, \sigma}}{\mathrm{d}\nu^2}
            -
            \frac{\mathrm{d}^2 \ln n \,\arg{\nu_0, 0}}{\mathrm{d}\nu^2}
        \right]
        \, .
    \label{eqn:A-of-tau-2nd-order}
\end{align}
At second order in $\Delta \nu$, there are two separate effects to consider:
\begin{enumerate}
    \item In the limit of low optical depth, the ratio of extinction to optical depth depends on the source spectrum.
    \item As optical depth increases, the nonlinear behavior of extinction with optical depth depends on the dust opacity spectrum.
\end{enumerate}
We can explore the second effect -- the non-linear dependence of extinction on dust column density -- by splitting $A\arg{\sigma}$ into a linear regime at low column density and nonlinear correction at larger optical depths:
\begin{align}
    A\arg{\sigma} &\simeq
        \frac{\mathrm{d}A \arg{\sigma=0}}{\mathrm{d}\sigma} \, \sigma
        +
        \Delta A \arg{\sigma} \, .
\end{align}
Generically, $\Delta A \propto A^2$ for any dust opacity law at $2^{\mathrm{nd}}$ order in $\Delta \nu$. At this order in $\Delta \nu$, the quadratic coefficient is determined by the behavior of the dust opacity relation, $\kappa \arg{\nu}$, with the shape of the source spectrum providing corrections at higher order in $\Delta \nu$. At $2^{\mathrm{nd}}$ order in $\Delta \nu$, the effect of the source spectrum is to alter the slope of the extinction-vs.-optical-depth relationship at low dust column density. 
\begin{align}
    \Delta A &\simeq
    -\frac{\ln 10}{5}
    \left[ \Delta \nu \, \frac{\mathrm{d} \ln \kappa \,\arg{\nu_0}}{\mathrm{d}\nu} \right]^{2}
    A^2
    \, .
\end{align}
With a power-law dust opacity, $\kappa \arg{\nu} \propto \nu^{\gamma}$, one obtains
\begin{align}
    \Delta A &\simeq
    -\frac{\ln 10}{5} \left( \frac{\Delta \nu}{\nu_0} \right)^{\!\!2} \gamma^2 A^2 \, .
    \label{eqn:DeltaA-of-gamma}
\end{align}
For small dust optical depths, extinction increases proportionally to optical depth in all passbands. However, as the dust column increases, the widest passbands deviate more quickly from this linear relationship than the narrower passbands, so that the ratio of extinction in different passbands is no longer constant. The vector $\vec{R}$ is no longer sufficient to determine the relationship between extinction in the different passbands.

We can estimate the scale of this effect for Gaia $G$ band. Here, we use the Gaia $G$ transmission curve calculated by \citet{ApellanizWeiler2018} and the \citet{Fitzpatrick1999} extinction curve, improved by \citet{Indebetouw2005} in the near-infrared \citep{SVO1,SVO2}. We set $\nu_0$ and $\Delta \nu$ to the transmission-weighted mean and standard deviation of frequency, respectively, and estimate $\gamma$ from the shape of the extinction curve between $\nu_0-\Delta \nu$ and $\nu_0+\Delta \nu$, obtaining $\nu_0 = 640\,\mathrm{nm}$, $\Delta \nu = 144\,\mathrm{nm}$ and $\gamma = 1.31$. Plugging these values into Eq.~\eqref{eqn:DeltaA-of-gamma}, we estimate $\Delta A_G \simeq -0.040 \, A_G^2$. This extinction non-linearity therefore becomes comparable to our photometric uncertainty floor of 0.02~mag at an extinction of $A_G \sim 0.7\,\mathrm{mag}$.

Numerically evaluating extinction using Eq.~\eqref{eqn:extinctinon-def} yields similar results. We define $\Delta A$ to be the difference between the calculated extinction at a given optical depth and the extinction that would be obtained by linearly extrapolating extinction from the low-optical-depth limit. With a black-body source spectrum at $6000\,\mathrm{K}$, we obtain $\Delta A_G = -0.037 \, \mathrm{mag}$ when $A_G = 1 \, \mathrm{mag}$, closely matching our $2^{\mathrm{nd}}$-order perturbative result. The numerical result only varies by 20\% for source temperatures between 4000~K and 8000~K.

\bibliography{paper}{}
\bibliographystyle{aasjournal}

\end{document}